\documentclass{emulateapj}
\usepackage{natbib}
\usepackage{rotating}
\newcommand{\tv}{$\tau_{v}$}
\newcommand{\td}{$\scriptstyle\sim$}

\begin{document}

\title{Young, UV-bright stars dominate dust heating in star forming galaxies}

\author{Ka-Hei Law}
\affil{Dept. of Physics and Astronomy, Johns Hopkins University, 3400 N. Charles St., Baltimore, MD 21218; klaw@pha.jhu.edu}

\author{Karl D. Gordon}
\affil{STScI, 3700 San Martin Dr., Baltimore, MD 21218; kgordon@stsci.edu}

\and
\author{K. A. Misselt}
\affil{Steward Observatory, University of Arizona, 933 N. Cherry Ave., Tucson, AZ 85721; misselt@as.arizona.edu}

\begin{abstract}
In star forming galaxies, dust plays a significant role in shaping the
ultraviolet (UV) through infrared (IR) spectrum.  Dust
attenuates the radiation from stars, and re-radiates the
energy through equilibrium and non-equilibrium emission.  Polycyclic
aromatic hydrocarbons (PAH), graphite, and silicates contribute to
different features in the spectral energy distribution; however, they
are all highly opaque in the same spectral region -- the UV.  Compared
to old stellar populations, young populations release a higher
fraction of their total luminosity
in the UV, making them a good source of the energetic UV photons that
can power dust emission.  However, given their relative abundance,
the question of whether young or old stellar populations provide most of these photons that power
the infrared emission is an interesting question.  Using three samples
of galaxies observed with the Spitzer Space Telescope and our dusty
radiative transfer model, we find that young stellar populations (on the order of
100 million years old) dominate the dust heating in star forming
galaxies, and old stellar populations (13 billion years old) generally
contribute less than 20\% of the far-IR luminosity. 
\end{abstract}


\section{INTRODUCTION}\label{intro}

The infrared radiation from most star-forming galaxies is dominated by
emission from dust grains heated by absorbed stellar energy.  Dust
emission is powered by absorption of radiation from ionizing and
non-ionizing stars.  Dust is most efficient at absorbing photons in
the ultraviolet (UV) as the relative optical depth of dust is the
highest in the UV \citep{Gordon2003}. Only early type (O and B) stars
produce significant amounts of UV photons; however, these hot massive
stars have short lifetimes (less than 100 million years) and are
formed in relatively small numbers compared to less massive, less
luminous, and cooler stars that produce very few UV photons.  Given
the initial mass function and evolutionary history, this implies that
star-forming galaxies have a small mass fraction of UV bright, young
stars as compared to UV faint, old stars.  Thus, the question arises:
which population of stars dominates the dust heating in star-forming
clouds?  The less numerous but much brighter in the UV young stars ($<
100$ Myr) or the numerous but much fainter in the UV old
stars?  How does this answer change when we consider the emission
at specific IR wavelengths?

The majority of the IR energy from star-forming galaxies is emitted at
far-IR (\td 100~\micron) wavelengths.  Historically, this far-IR
emission has been identified as infrared cirrus emission from dust
heated by non-ionizing populations \citep{Helou1994},
which are older than \td 10 Myr.
\citet{Cirrus1987} interpreted the far-IR emission from spiral disks
in terms of two thermal components with different temperatures and found that
the cirrus component contributes more than half of the total far-IR flux.
However, as old
stars emit very few of the UV photons that power dust emission, it is
possible that UV-bright young stars could dominate cirrus emission.
For example, a small number of young stars embedded in a large
optically thin cloud can result in a dilute radiation field, cold dust
temperature and therefore cold cirrus emission.

Star formation rate (SFR) indicators are important observational
probes of the star formation histories of galaxies. They are
usually single-band or wavelength-integrated quantities that are
presumed to trace a specific regime of recent star formation in a
region or galaxy \citep{Kennicutt1998}. The most common SFR
indicators include the H$\alpha$ flux (tracing unobscured ionizing
stars, $< 10$ Myr), UV (tracing unobscured ionizing and UV-bright,
non-ionizing stars, $< 100$ Myr), and total infrared (TIR, tracing
obscured star formation).

The UV and H$\alpha$ flux are heavily attenuated by dust,
with a typical extinction of 0-4 mag and 0-2 mag respectively \citep{Kennicutt2009}.
Since the UV and H$\alpha$ flux only trace the stellar light unabsorbed by dust,
an accurate estimation of star formation activity requires a correction factor
to account for the effect of dust. For starburst galaxies,
emission-line diagnostics and UV colors
may be used to such purpose, but they are often difficult to obtain or highly
uncertain \citep{Kennicutt2009}. An alternative way is to look at the IR flux,
which accounts for the energy missing in the UV and the optical \citep{Calzetti2007}.
\citet{Kennicutt2009} found that the combination of H$\alpha$ and TIR provides
a robust SFR measurement. \citet{Leroy2008} and \citet{Bigiel2008} used
the 24 \micron\ flux combined with the GALEX far-UV to study the SFR in nearby galaxies.

TIR alone as a SFR indicator suffers from a number of problems.
It does not trace unobscured star formation, and for the typical amount
of dust in galaxies, a non-trivial amount of energy
escapes in the UV and optical. Recalling the ``cirrus emission" problem, TIR
can only work well as an SFR indicator if
the total infrared emission correlates well with newly formed stars
($< 100$~Myr), instead of old stars that formed long ago.  Therefore,
understanding whether young stars or old stars dominate the TIR dust
emission is an important step in interpreting the TIR SFR indicator.
In addition, TIR is often integrated over a sparsely sampled
wavelength interval in
intermediate or high-redshift galaxies,
which may introduce many uncertainties \citep{Calzetti2007}.
\citet{Sauvage1992} found a systematic decrease of the L(far-IR)/L(H-alpha)
ratio from early- to late-type spirals, and suggested a systematically
varying cirrus fraction
as the explanation, accounting for 86 percent of L(far-IR) for Sa galaxies
to about 3 percent for Sdm galaxies.
\citet{Xu1996} also found that the heating of the diffuse dust is
dominated by optical radiation from stars at least a billion years old in M31.
With these results, it would be important to subtract the cirrus emission
from the total infrared luminosity in the calculation of star formation rate.

In the simplest model, dust emission can be thought of as modified blackbody
radiation at a certain temperature plus emission features.  However,
equilibrium dust emission is not the only possible emission path.
\citet{Duley1973} pointed out that non-equilibrium dust emission should
also be important and this has been confirmed observationally
\citep[e.g.][]{Sellgren1984}.  Upon the absorption of an
energetic photon, a large molecule or small dust grain can
attain a very high temperature for a short period of time before it
cools.  In many situations, if non-equilibrium heating is not
considered, the observed mid-IR luminosity implies an
unrealistically high equilibrium dust temperature.  While
older stars can rarely excite dust grains to produce non-equilibrium
heating, this type of heating clearly dominates the mid-IR emission
from star-forming galaxies.

Recent interesting results from the Herschel observatory have begun
exploring the problem of the source of dust heating using the far-IR
and sub-mm. Specifically, in a study of M81, \citet{Bendo2010} found that the 
far-IR to sub-mm (160-500 \micron) emission is dominated by dust heated by evolved
stars while the 70 \micron\ emission is caused by the active galactic nucleus 
and young stars in star forming regions.
From a study of 51 nearby galaxies, \citet{Boselli2010} found that the
warm dust (f60/f100) correlates with star formation, while cold dust (f350/f500)
anti-correlates. Our study does not probe cold dust in the submillimeter
range, but we plan to do so in a later paper.

In this paper, we examine the relationship between IR luminosity and the
age of stellar populations.  Through the use of a self-consistent radiative transfer
and dust emission model, observed dust properties, and stellar and dust 
geometries designed to represent galactic environments, we compare our results to
observed galaxies (the SINGS sample, \citet{SINGS}, the starburst galaxies in
\citet{Engelbracht2008}, and the LVL sample, \citet{LVL}) and attempt
to give a quantitative answer on the age of the stellar populations that dominates
the IR emission in star forming galaxies.

\section{METHOD}\label{method}
\subsection{Model}\label{model}

\begin{figure*}
	\plottwo{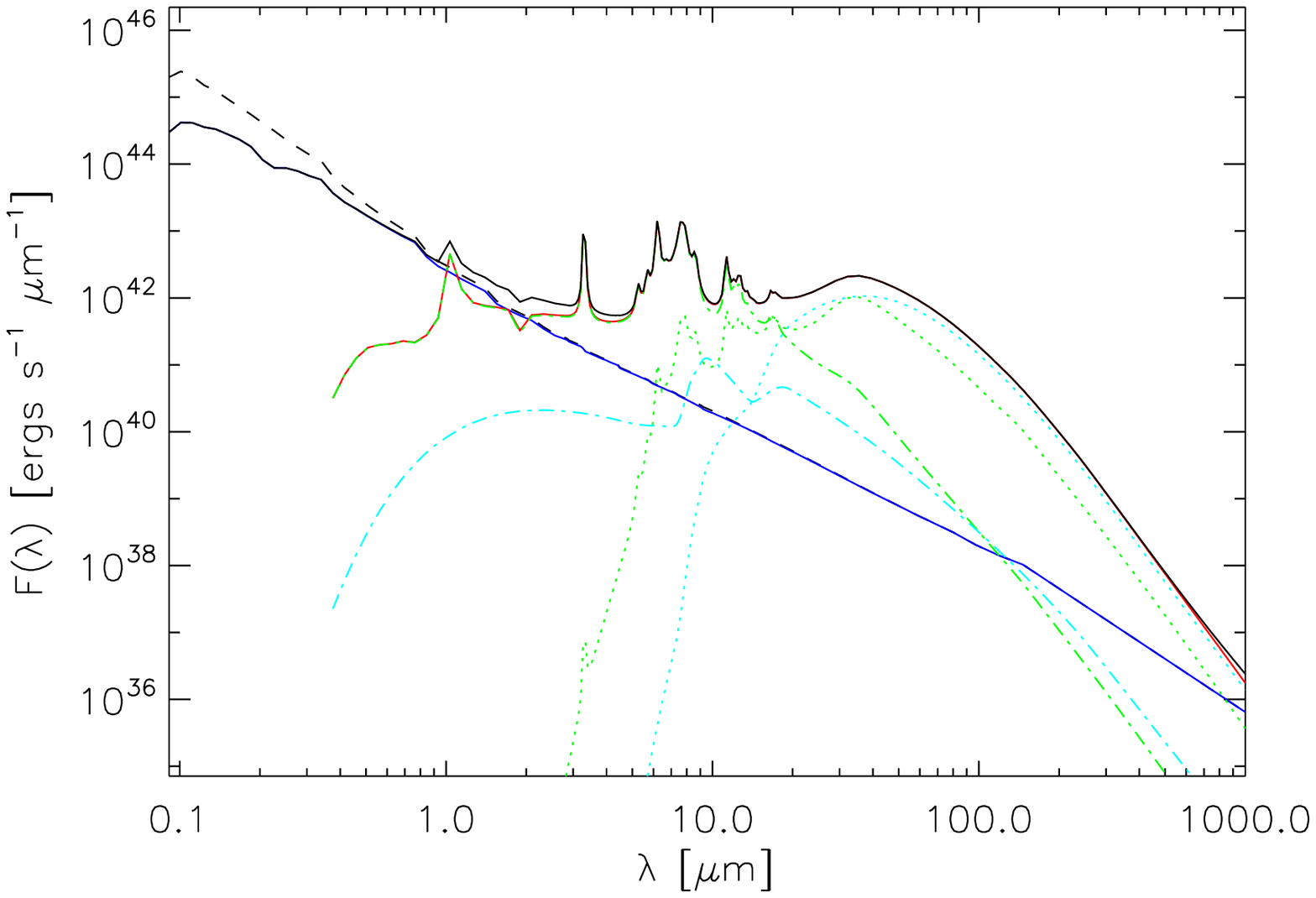}{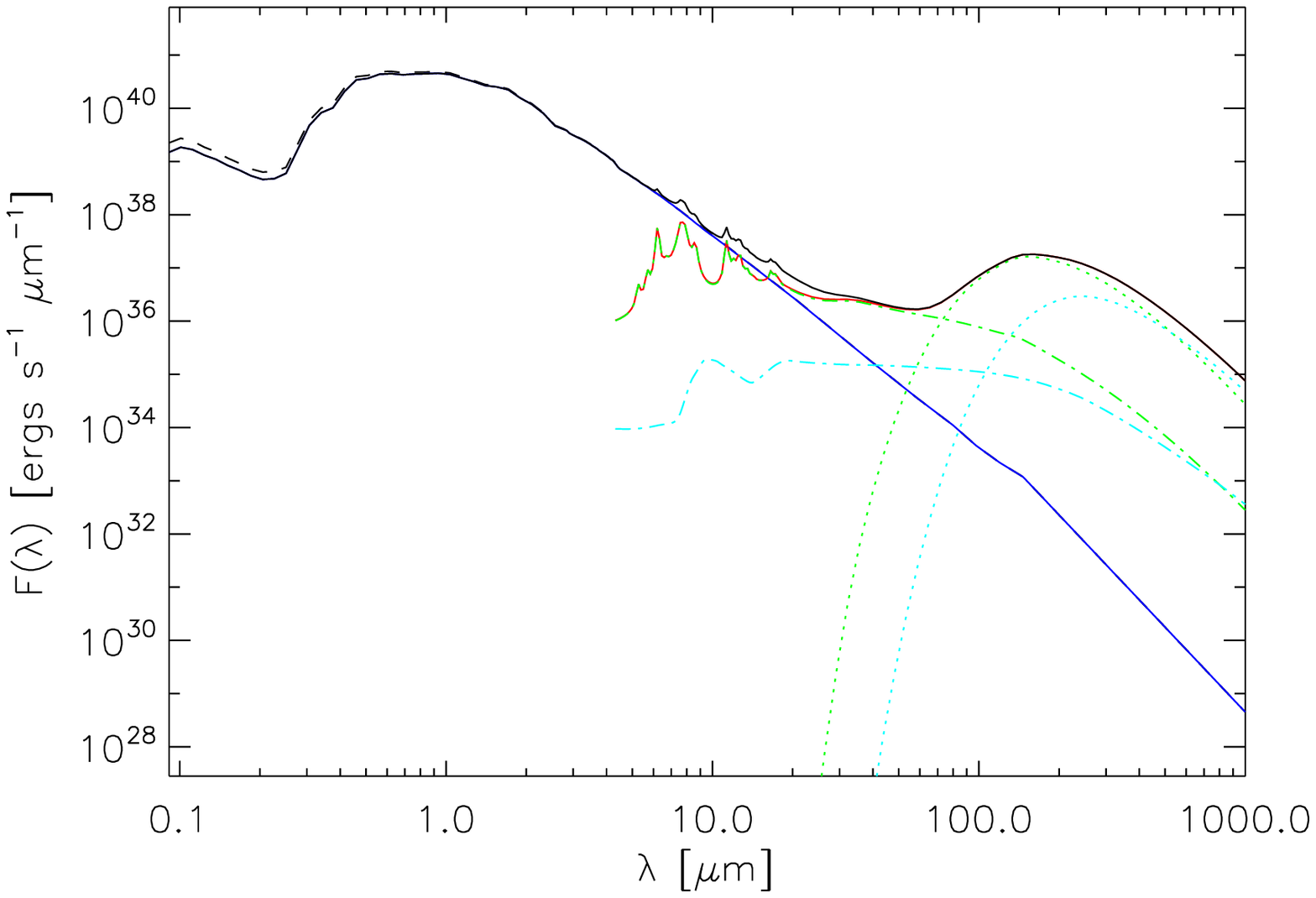}
	\caption{
		Example global SED outputs from DIRTY for a young (10 Myr, left) and
		an old (13 Gyr, right) stellar populations, each with
		a mass of $10^{10}$~M$_{\sun}$, Milky Way type dust, 10 kpc radius, and an optical depth
		of $\tau_{v} = 1$.
		The former has a SHELL geometry and the latter has a CLOUDY geometry.
		The dashed black line is the input stellar SED.
		The solid black, blue and red lines are the total output SED, the radiative
		transfer component (diffuse/extincted plus scattered) and the total dust emission.
		Among the dust emission components, the green and light blue
		lines represent carbonaceous and silicate grains; the dotted and
		dash-dotted lines represent equilibrium and non-equilibrium emission, respectively.
	}
	\label{fig:dirtysed}
\end{figure*}

DIRTY (DustI Radiative Transfer, Yeah!) is a self-consistent Monte
Carlo radiative transfer model \citep{DIRTY1, DIRTY2}.  Due to its Monte
Carlo nature, it allows for arbitrary dust and stellar distributions.
It computes emission from the three standard dust grain components
\citep{Weingartner2001} - carbon grains, silicates, and polycyclic
aromatic hydrocarbons (PAH).  One of the key strengths of DIRTY is
that it is completely self-consistent.  It avoids assumptions whenever
possible; the radiation field is directly calculated from radiative
transfer, instead of the usual method of scaling the standard solar
neighborhood radiation field \citep{Draine2009}.  
The initial radiation field is obtained from an external spectral
evolutionary synthesis model.  Both equilibrium and
non-equilibrium emission are calculated from the radiation field and
their contributions to the radiation field itself are iteratively taken
into account.  DIRTY has been used to model a variety of situations, for
example, the dusty starburst nucleus of M33 \citep{Gordon1999} and the
general behavior of galaxies \citep{Witt2000}.  Figure~\ref{fig:dirtysed}
shows examples of global spectral energy distributions (SED) from DIRTY.

\begin{table*}
	\caption{Model parameters used in DIRTY.}
	\centering
	\begin{tabular}{c c}
		\hline \hline
		Parameter & Values \\
		\hline
		Optical depth at V band ($\tau_{v}$) & 0.2 - 5.0 \\
		Stellar age (Myr) & 1 - 13000 \\
		Radius (pc) & 100 - 10000 \\
		Dust model & Milky Way dust (with $R_V = 3.1$, $b_C = 6 \times 10^{-5}$) and SMC Bar dust from \citet{Weingartner2001} \\
		Stellar model & PEGASE 2 \citep{PEGASE} with instantaneous bursts \\
		Metallicity & Solar (0.02) and 1/5 Solar (0.004) \\
		\hline
	\end{tabular}
	\label{table:params}
\end{table*}

Young and old galaxies are best described by different dust geometries,
which can lead to different dust absorption efficiencies and effective
optical depths \citep{Witt1992}.  Following the notation of \citet{Witt2000},
we use the CLOUDY and SHELL geometries.  In the CLOUDY geometry, stars
are uniformly distributed in a spherical volume, and a dusty core is
located within 0.69 times of the radius of the stellar distribution;
whereas in the SHELL geometry, stars extend only to 0.3 times of the
outer radius and are surrounded by a concentric shell of dust
from 0.3 to 1.0 of the outer radius. See Figure 1 of \citet{Witt2000} for a
graphical description of the CLOUDY and SHELL geometries. 

For both geometries, the stellar density per unit volume is constant.
The CLOUDY model reproduces the general characteristics
of old stellar populations in galaxies with strong central bulges,
where the majority of stars are found outside of the dust.  The SHELL
model mimics a young star cluster with surrounding clouds.
It has been shown that starbursts require a SHELL geometry
to explain various color-color plots \citep{Gordon1997} and the
attenuated-to-intrinsic ratios of hydrogen lines \citep{Calzetti2001}.
In this study, we use SHELL for stars that are 100 Myr old or younger and
CLOUDY for older ones.  Other model parameters can be found in
Table~\ref{table:params}.

\citet{Witt1996} found that it is necessary to model
interstellar dust as a multi-phase medium.  To achieve this, a
small fraction (15 \% total filling factor) of clumps of high density
dust is randomly placed into a field of low density dust with 100
times lower density in the global dust geometries mentioned previously.
Without the clumpiness, the effectiveness of dust
absorption would be overestimated.  In this study, we exclusively use
clumpy dust distributions.
The optical depth averaged over all sightlines (from infinity
to the center of the geometry) is normalized to the desired value of \tv.

We use the spectral evolutionary synthesis (SES) model PEGASE 2 \citep{PEGASE}
as the stellar input to our model.  We model starbursts of various ages with
solar and 1/5 solar metallicity and the Padova evolutionary tracks.
An old galaxy may contain both old stellar populations and young stellar populations,
although the latter is expected to be much less abundant considering that the
typical timescale for gas depletion is about 3 Gyr \citep{Pflamm2009}.
The time-dependent profile of star formation could be modeled as an
exponentially decaying burst \citep{exp_time1973,exp_time2003}.
However, since we want to model the characteristics (the luminosity ratios)
of stellar populations at a certain age, we use instantaneous starbursts.
To bracket the possible real world scenarios, we model the extreme cases
of very young (1 Myr) and very old (13 Gyr) starburst populations.
Gas continuum and emission of recombination lines, as calculated by PEGASE,
are included.
See \citet{Gordon1999} for a detailed discussion on how we use spectral
evolutionary synthesis models with our dusty radiative transfer model.

The dust extinction curve for stars in the Milky Way (MW), Small and Large
Magellanic Clouds (SMC and LMC) are found to have overall similar shapes
with significant variation in the UV \citep{Gordon2003}.  There are
two distinct features that differentiate the different types of dust.
The first one is the 2175~\AA\ bump, which is found to vary in strength
on average between the MW diffuse ISM/LMC general (strong bump), LMC2
(near 30 Dor)/SMC Wing (weak bump), and SMC Bar (no bump).  The second is
the far UV rise, which generally varies in strength inversely with the 2175~\AA\
bump.  Dust properties can be influenced by star forming activity and
metallicity.  In particular, \citet{Gordon1997} found that
the SMC Bar type dust (lacking a 2175~\AA\ bump) describes the starburst
galaxies better than either the LMC or MW type dust.
From the GMASS survey, \citet{Noll2009} found that
there is a wide range of UV dust properties, including those that are
intermediate between SMC Bar and LMC2 type dusts.  Modeling results
show that the type of dust can affect the strength of PAH features
\citep{Draine2007}, which in turn affects the IR observations.  As a
result, it is important to choose an appropriate dust type for our study,
and to explore the sensitivity of our results to the dust type.
To span the whole range of known dust properties, we use SMC Bar type dust
and Milky Way type dust.  For a fair comparison,
we use the dust models of \citet{Weingartner2001} for both the SMC and
Milky Way type dust.

Radius and stellar mass are not independent dimensions in our model.
Since the radiation intensity drops as the square of the radius, an
increase in the model radius by a factor of $x$ can be compensated by
an increase in the stellar mass by a factor of $x^{2}$.  The two cases
give the same radiation intensity in each grid cell in the model, and
therefore the same dust temperature.  The resultant SEDs will have the
same shape, with the only difference being the lower overall luminosity
in the latter case.  This has been confirmed by test runs of our model;
for example, a stellar population with $10^{11}$ solar masses and 10 kpc
radius gives the same SED as the one with $10^9$ solar masses and 1 kpc
radius, when the luminosity in the latter is scaled up by a factor of 100
(all the other parameters are unchanged).  Since our study is based on
the ratios of luminosities, a larger, more luminous galaxy and a smaller,
dimmer galaxy with same radiation field intensity give the same result.

Due to the Monte Carlo and iterative nature of our model,
the run time varies significantly depending on the parameters.
The fastest models take about 4 CPU hours on a 2.33 GHz Intel Xeon processor,
while the slowest ones often need about 2 days.
The average run time is about 20 hours.
Each model uses a single thread and we launch multiple processes in
parallel to utilize all the CPU cores in multi-core processors. 
We iterate the radiative transfer and dust emission processes until
the global energy conservation error is within 5\%. 
The resultant uncertainty of flux at each wavelength is usually within 1\%.

\subsection{Luminosity Ratios}\label{lratio}

To compare the modeled SED with infrared observations, we calculate the
luminosities in the IRAC and MIPS bands with calibrated response
functions \citep{IRACresp,MIPS24,MIPS70,MIPS160}.
Since the galaxies in our sample have a very wide range of total luminosity,
we study luminosity ratios instead of raw luminosities.
We normalize the mid- and far-IR luminosities by the IRAC1 (3.6 \micron) luminosity,
and the resulting ratios are the characteristics of the dust and stellar populations
that we compare with our model.
These luminosity ratios are measures of the efficiency of the stellar and dust
distribution at producing radiation in a certain wavelength regime
(relative to IRAC1).
At 3.6 \micron, the IRAC1 band is only mildly contaminated by dust
emission (e.g., the 3.3~\micron\ aromatic/PAH feature),
and is less affected by dust extinction than the shorter wavelengths.
The IRAC1 luminosity can be considered as a proxy of stellar luminosity and an approximate measure of mass,
although the mass-to-light ratio depends on their underlying stellar populations and the infrared colors \citep{Bell2001}.
The combination of the PEGASE SES and DIRTY models account
for stellar age, dust emission, and absorption self-consistently.
As we see in later discussions, the luminosity ratios are robust
against changes in radius or total stellar luminosity.

\subsection{Data}\label{data}

\begin{figure}
	\plotone{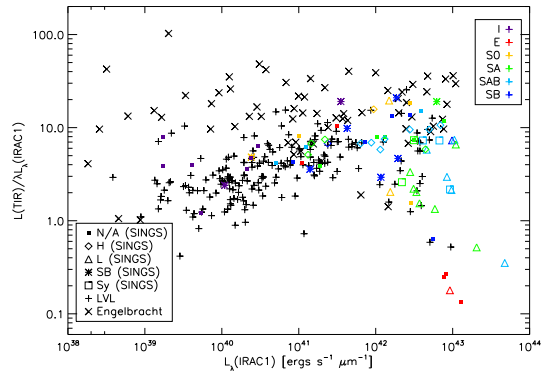}
	\caption{
		TIR to IRAC1 luminosity ratio vs IRAC1 luminosity for our sample.
		For the SINGS galaxies \citep{SINGS}, morphological types are shown
		in different colors and nuclear types are shown in different symbols.
		The SINGS galaxies span a large range of galaxy types, from spirals (S) to
		ellipticals (E) and irregulars (I).
		The nuclear type H, L, SB and Sy stand for H II,
		LINER, starburst, and Seyfert. The \citet{Engelbracht2008} starburst
		galaxies and the LVL galaxies \citep{LVL} are shown in black ``$\times$" and ``+" respectively. 
	}
	\label{fig:sings}
\end{figure}

From the SINGS (Spitzer Nearby Galaxies Survey) dataset \citep{SINGS},
we have good measurements of the UV, visible and IR fluxes for a number
of nearby galaxies.  They range from spiral to elliptical to irregular
galaxies, and are a good representation of a wide range of galaxies.
We removed 4 galaxies (M81 Dwarf A, NGC 3034, Holmberg IX, and DDO 154)
for which the IRAC or MIPS band fluxes are not well measured (either an
upper limit or saturated).
For the remaining 71 galaxies, we calculate the luminosity from the flux data
from \citet{SINGSData} and distance data from \citet{SINGS}.  
The metallicity is calculated from the average of the ``high" and ``low"
oxygen abundance values from \citet{Moustakas2010}; the same method
was used in \citet{Calzetti2010}.
Figure \ref{fig:sings} shows the wide range of galaxies types in the SINGS
sample, where we have used Eq. 22 from \citet{Draine2007b} to compute
$L_{TIR}$.  

To expand the range of galaxies studied, we include the starburst galaxies
from \citet{Engelbracht2008} and the Local Volume Legacy (LVL) sample from \citet{LVL}.
The starburst galaxies have a higher fraction of recent star formation
and UV-bright young stars.  Their data include imaging and spectroscopy from
the Spitzer Space Telescope, as well as ground-based near-infrared imaging.
The Engelbracht sample consists of 66 local star forming galaxies of which 65
have high quality IRAC and MIPS data available; UM~420 was dropped from our
sample because the MIPS data were not available.
For this sample we take all flux, distance and metallicity values
from \citet{Engelbracht2008}.  On the other hand, the LVL is a statistically
unbiased sample of 258 galaxies in the local universe out to 11 Mpc which
consists mainly of dwarf galaxies, and
we take flux and distance values from \citet{LVL} and metallicity values
from \citet{Marble2010}.
We removed galaxies without good IRAC and/or MIPS measurements (no data
or flux available only as an upper bound) and galaxies without
metallicity data, and arrived at a sample of 194 galaxies.

\begin{table*}
	\caption{
		Statistics of the luminosity ratios ($L_\lambda(\lambda_1)/L_\lambda(\lambda_2)$).
	}
	\centering
	\begin{tabular}{lc cccccc}
		\hline \hline
		Dataset & Count & IRAC2/IRAC1 & IRAC3/IRAC1 & IRAC4/IRAC1 & MIPS24/IRAC1 & MIPS70/IRAC1 & MIPS160/IRAC1 \\
		\hline
		SINGS & 71 & 0.433 $\pm$ 0.073 & 0.411 $\pm$ 0.274 & 0.438 $\pm$ 0.362 & 0.110 $\pm$ 0.151 & 0.131 $\pm$ 0.122 & 0.049 $\pm$ 0.031 \\
		\cline{2-8}
		~~~H II nuclei & 13 & 0.469 $\pm$ 0.147 & 0.567 $\pm$ 0.402 & 0.604 $\pm$ 0.329 & 0.140 $\pm$ 0.181 & 0.141 $\pm$ 0.058 & 0.066 $\pm$ 0.017 \\
		~~~Starburst nuclei & 9 & 0.441 $\pm$ 0.036 & 0.499 $\pm$ 0.314 & 0.562 $\pm$ 0.457 & 0.249 $\pm$ 0.273 & 0.218 $\pm$ 0.181 & 0.053 $\pm$ 0.032 \\
		~~~LINER/Seyfert nuclei & 21 & 0.408 $\pm$ 0.029 & 0.317 $\pm$ 0.127 & 0.307 $\pm$ 0.211 & 0.055 $\pm$ 0.077 & 0.086 $\pm$ 0.128 & 0.037 $\pm$ 0.028 \\
		~~~Others & 28 & 0.434 $\pm$ 0.045 & 0.381 $\pm$ 0.244 & 0.419 $\pm$ 0.406 & 0.091 $\pm$ 0.093 & 0.133 $\pm$ 0.105 & 0.049 $\pm$ 0.034 \\
		\cline{2-8}
		~~~Type I & 9 & 0.465 $\pm$ 0.041 & 0.282 $\pm$ 0.186 & 0.226 $\pm$ 0.288 & 0.138 $\pm$ 0.224 & 0.145 $\pm$ 0.121 & 0.035 $\pm$ 0.019 \\
		~~~Type E & 6 & 0.397 $\pm$ 0.034 & 0.232 $\pm$ 0.163 & 0.188 $\pm$ 0.271 & 0.050 $\pm$ 0.095 & 0.061 $\pm$ 0.104 & 0.014 $\pm$ 0.021 \\
		~~~Type S0 & 7 & 0.507 $\pm$ 0.200 & 0.616 $\pm$ 0.622 & 0.594 $\pm$ 0.630 & 0.245 $\pm$ 0.262 & 0.247 $\pm$ 0.207 & 0.047 $\pm$ 0.036 \\
		~~~Type SA & 19 & 0.420 $\pm$ 0.035 & 0.418 $\pm$ 0.208 & 0.451 $\pm$ 0.325 & 0.074 $\pm$ 0.116 & 0.100 $\pm$ 0.087 & 0.053 $\pm$ 0.031 \\
		~~~Type SAB & 18 & 0.418 $\pm$ 0.023 & 0.434 $\pm$ 0.150 & 0.515 $\pm$ 0.294 & 0.082 $\pm$ 0.064 & 0.113 $\pm$ 0.072 & 0.059 $\pm$ 0.026 \\
		~~~Type SB & 12 & 0.431 $\pm$ 0.041 & 0.431 $\pm$ 0.255 & 0.495 $\pm$ 0.342 & 0.137 $\pm$ 0.143 & 0.164 $\pm$ 0.143 & 0.059 $\pm$ 0.033 \\
		\hline
		Engelbracht & 65 & 0.518 $\pm$ 0.184 & 0.578 $\pm$ 0.395 & 0.738 $\pm$ 0.632 & 0.636 $\pm$ 0.612 & 0.439 $\pm$ 0.345 & 0.094 $\pm$ 0.163 \\
		\hline
		LVL & 194 & 0.436 $\pm$ 0.037 & 0.287 $\pm$ 0.158 & 0.237 $\pm$ 0.225 & 0.061 $\pm$ 0.104 & 0.097 $\pm$ 0.076 & 0.033 $\pm$ 0.022 \\
		\hline
	\end{tabular}
	\label{table:stat}
\end{table*}

\begin{figure*}
	\plottwo{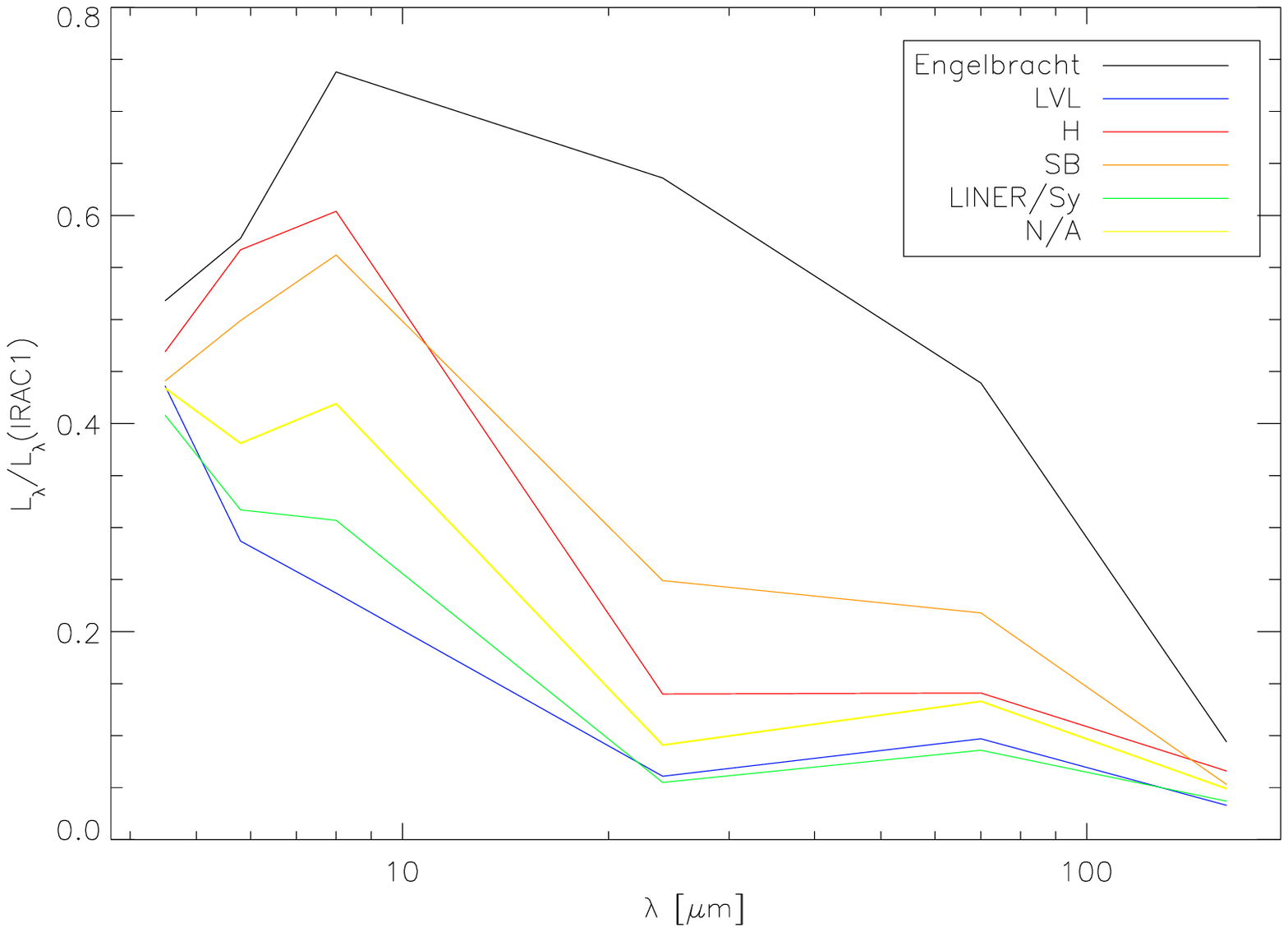}{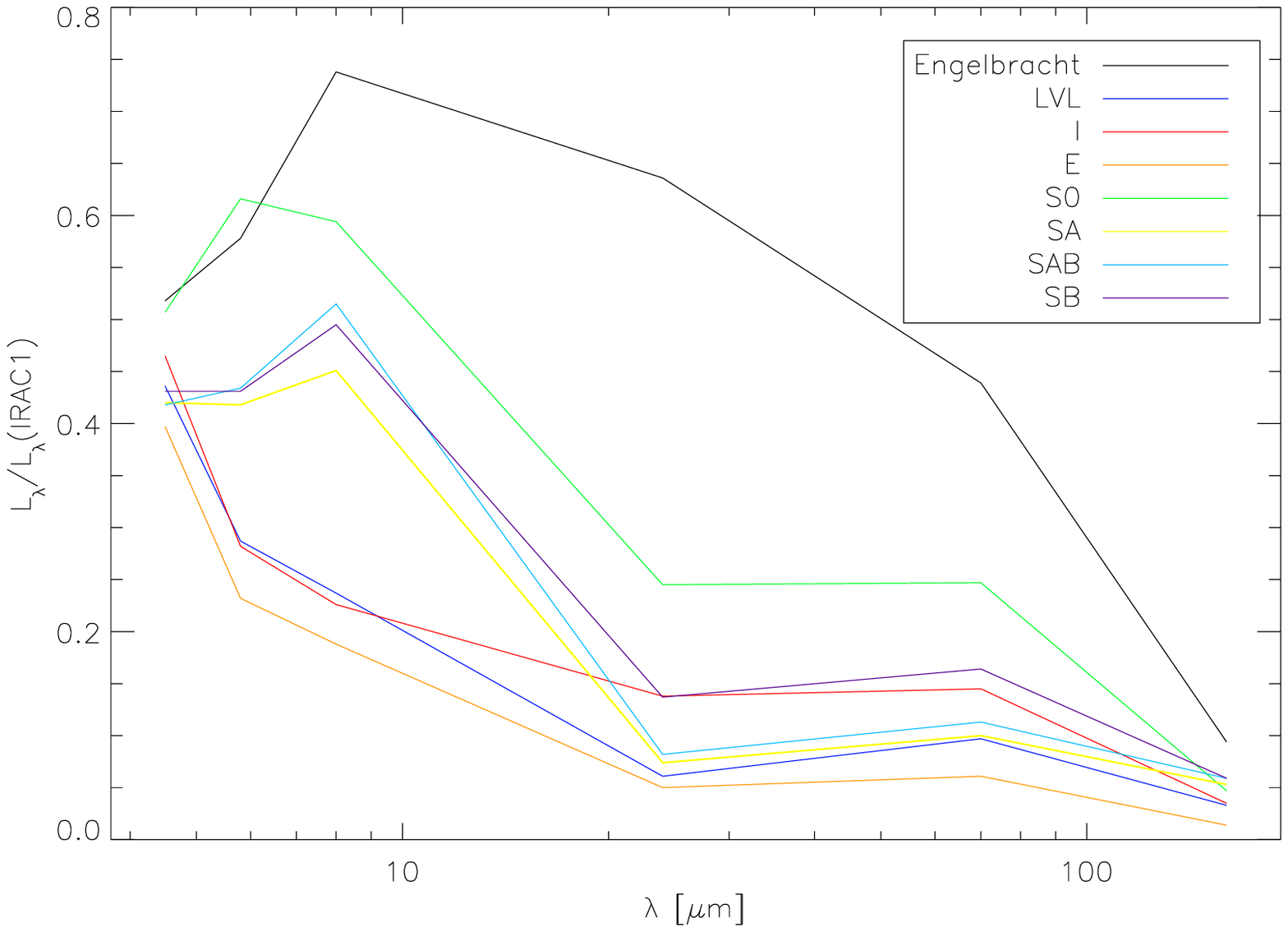}
	\caption{
		The average SEDs (normalized by IRAC1 luminosity) of galaxies with
		different nuclear types (left) and morphological types (right), using the
		data in Table~\ref{table:stat}.
		Note that the \citet{Engelbracht2008} starburst galaxies show a
		significantly higher ratio in MIPS24 and MIPS70 compared to the
		other galaxies.
		On the other hand, elliptical galaxies (orange, right) give the lowest
		luminosity ratios, as they are the least efficient in producing
		mid- and far-IR fluxes.
	}
	\label{fig:stat}
\end{figure*}

In Table~\ref{table:stat} we list the statistics of the band ratios
of our sample.  The SINGS galaxies are classified into different nuclei
and morphological types according to \citet{SINGS}.  We keep the Engelbracht
sample (as a group of starburst galaxies) and the LVL sample
(representing the dwarf galaxies) separate from the SINGS' categories
for a clear statistical comparison.
Each number in the table is the average (plus or minus
the standard deviation) of the ratio of the given band luminosity to
the IRAC1 luminosity.  Figure~\ref{fig:stat}
shows the table entries graphically.  Among the SINGS galaxies, the
starburst and H II nuclei tend to have higher luminosity ratios while those
without active nuclei tend to  have lower luminosity ratios.  Galaxies with
Seyferts (Sy) and LINERS (L) nuclei are in between the two extremes.
While our model does not simulate Seyfert (Sy) or LINERS (L) nuclei,
the luminosity ratios of these galaxies are within the range of the
other galaxies, and we keep them in our study.  Focusing on the MIPS160
column, we notice that the luminosity ratio for elliptical galaxies is
significantly lower than the other types of galaxies.  On the other hand,
the \citet{Engelbracht2008} starburst galaxies show a much higher ratio
in all the MIPS bands, meaning that they are more efficient in producing
far-infrared emission.

\section{RESULTS}\label{results}
\subsection{Dust Type}\label{dust}

\begin{figure*}
	\plottwo{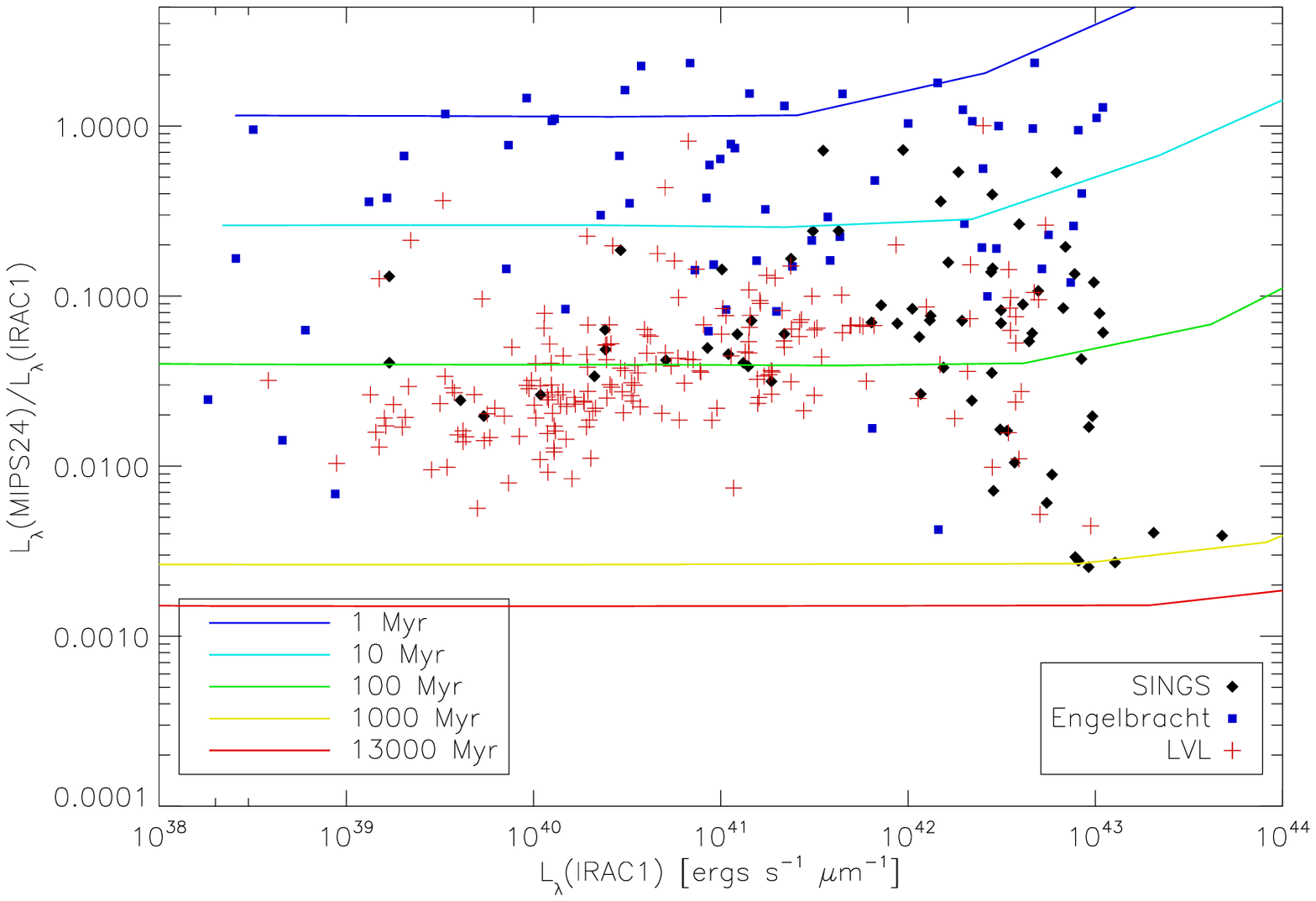}{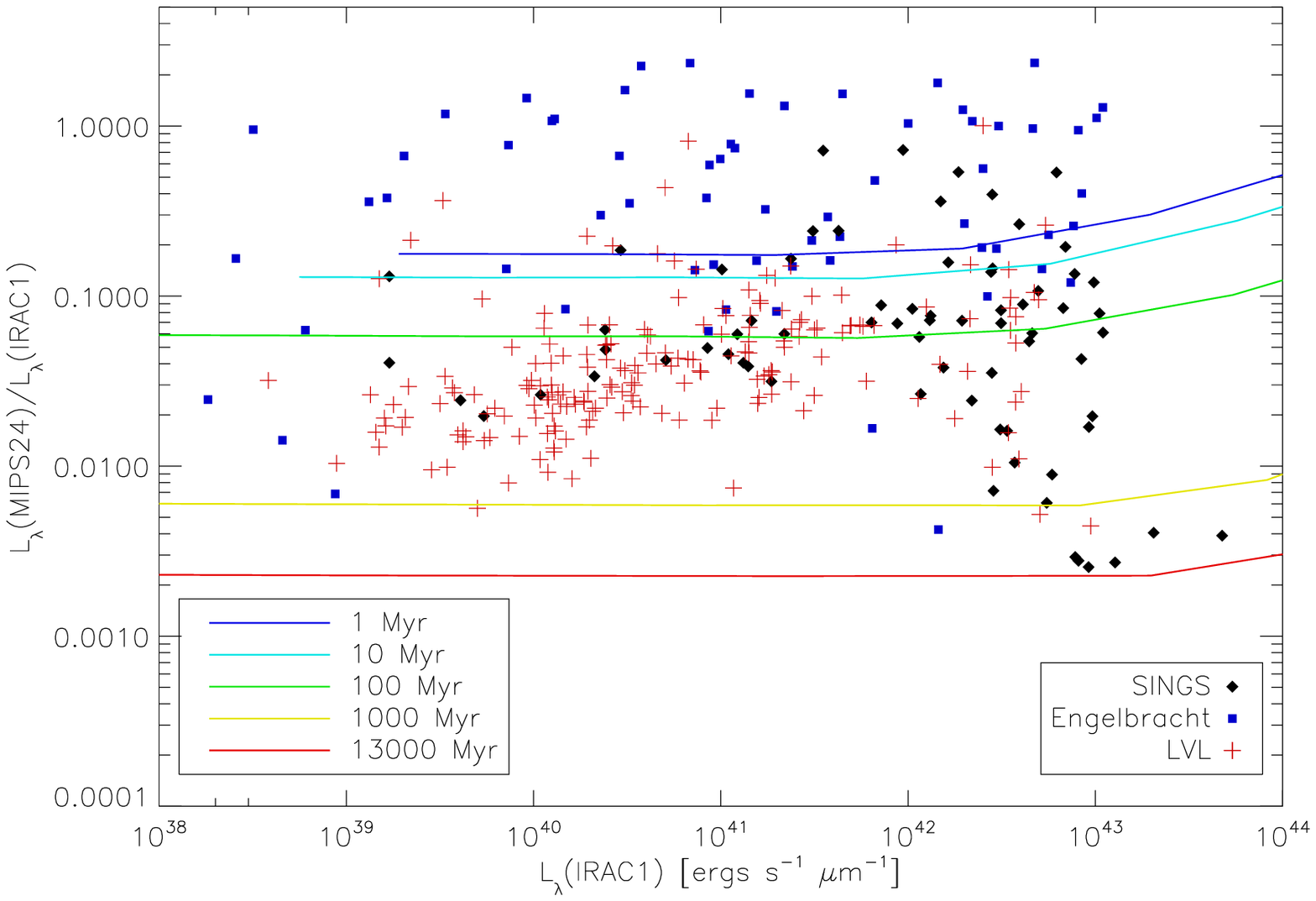}
	\caption{
		Comparison of SMC Bar (left) and Milky Way (right) type dusts.
		Shown in the figures are the MIPS24 (24 \micron) to IRAC1 (3.6 \micron) luminosity ratios against the IRAC1 luminosity.
		The curves represent results from the DIRTY radiative transfer models with different stellar ages,
		while the data points represent our galaxy sample.
	}
	\label{fig:dust_type}
\end{figure*}

In Figure~\ref{fig:dust_type}, we plot the MIPS24 to IRAC1 luminosity
ratio for our galaxy sample,
together with models with SMC Bar (left) and Milky Way type dust
(right).  Different models on the same curve have different stellar
mass, characterized by their different IRAC1 luminosity on the x-axis.
Here we use $\tau_{v} = 1.0$ and radius = 10 kpc, but we see similar
trends with other parameters.  We note that the curves are relatively
flat until they tick up in the regime of very high IRAC1 luminosity.
This is because dust emission in the MIPS24 band is dominated
by non-equilibrium heating for young stellar populations, or stellar
continuum for old stellar populations.  In either case, the fraction of
MIPS24 to IRAC1 is a constant.  Only at very high luminosity, does the dust
temperature become high enough for equilibrium heating to make a comparable
contribution and the curves start to turn up.  We discuss the
effects of non-equilibrium emission in section \ref{noneq}.

From 1 Myr old to 13 Gyr old, the models
with SMC Bar type dust roughly span the whole range of luminosity ratios
for the galaxies.  On the other hand, the models with MW
type dust are unable to cover all the galaxies.  Even at the youngest age
of 1 Myr, the model curve is below some galaxies (including
most of the \citet{Engelbracht2008} starburst galaxies).  At the other
extreme, both types of dust do equally well for low MIPS24 to IRAC1
luminosity ratios.  None of the galaxies have luminosity ratios
lower than our oldest models (13 Gyr, close to the age of the universe).
We see a similar distinction in the MIPS70 to IRAC1 luminosity ratios.
From this perspective, the SMC Bar type dust is a better choice for
our study, although this is different from~\citet{Draine2007} who
found that the SINGS galaxy sample has similar dust-to-gas ratio and
similar PAH abundance to MW type dust.  It is possible that the different
results from the dust types are due to dust processing. If the MW type dust
is more susceptible to destruction in a UV radiation field compared to the SMC Bar type dust,
the former will produce less far-IR emission per unit IRAC1 luminosity and
therefore a lower luminosity ratio.  Unless otherwise specified,
discussions in the following sections refer to SMC Bar type dust.
Despite the differences we discuss here, the conclusion we draw
about the age of stellar populations heating the dust (see section \ref{constrain})
is independent of the type of dust. 

\subsection{Age}\label{age}

To study what stellar age would best reproduce the observed luminosity
ratios, we fix the other model parameters and see how the model
results change as a function of stellar age.  Knowing that $\tau_{v}$
is on the order of unity for normal disk galaxies \citep{Holwerda2007},
we assume an optical depth of $\tau_{v} = 1.0$.
\citet{Dale2006} also found that the average attenuation
is $A_v = 1.0$ (which equals $\tau_{v} = 0.92$) for a large portion
of SINGS and some archival sources from ISO and Spitzer.
We set the radius to be 10 kpc, a reasonable size of a galaxy.  Our results do
depend on the choice of optical depth and radius, but we first see what
we find with these values.  In section \ref{tvradius} we examine the effect
of varying the optical depth and radius.

\begin{figure*}
	\plottwo{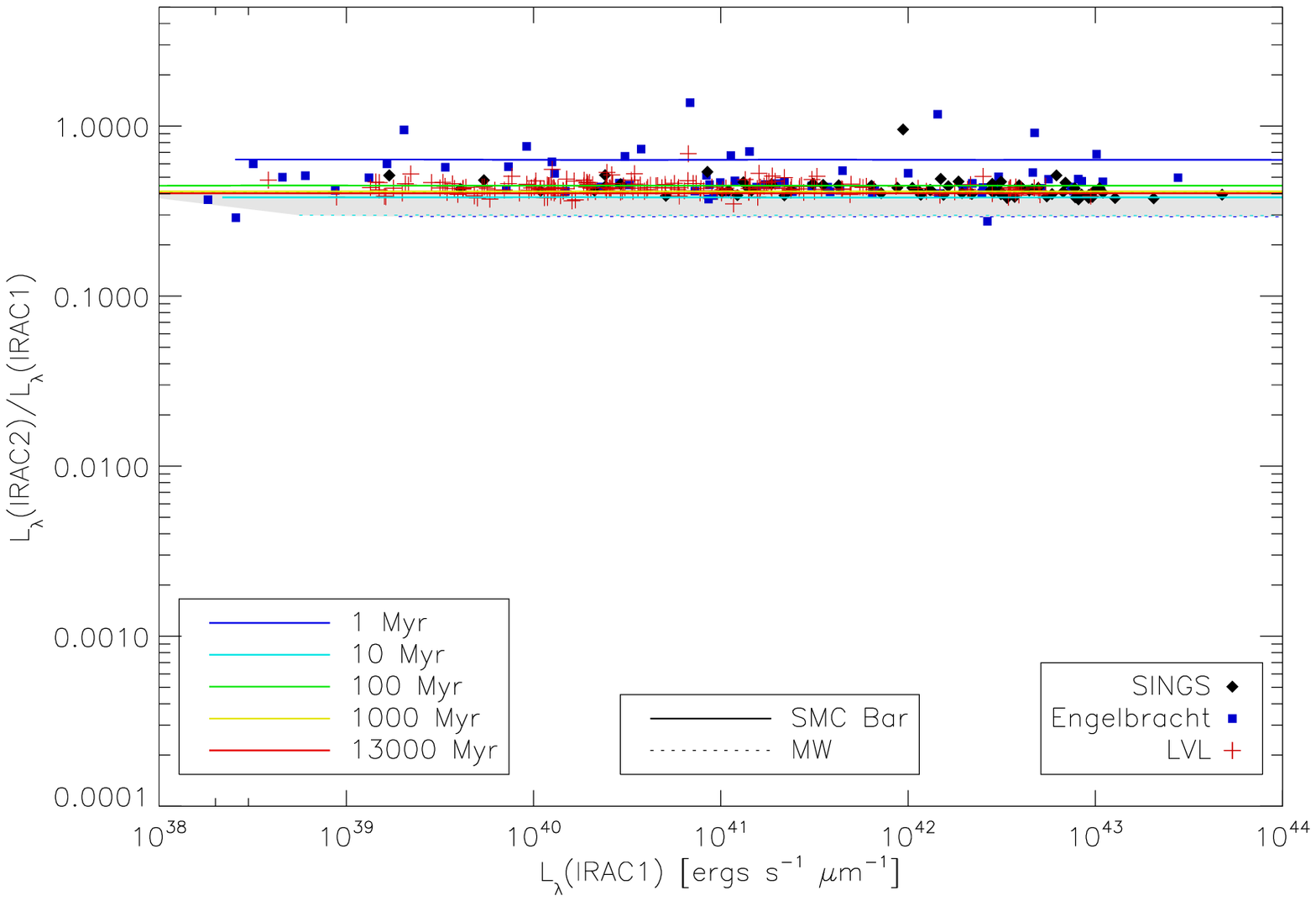}{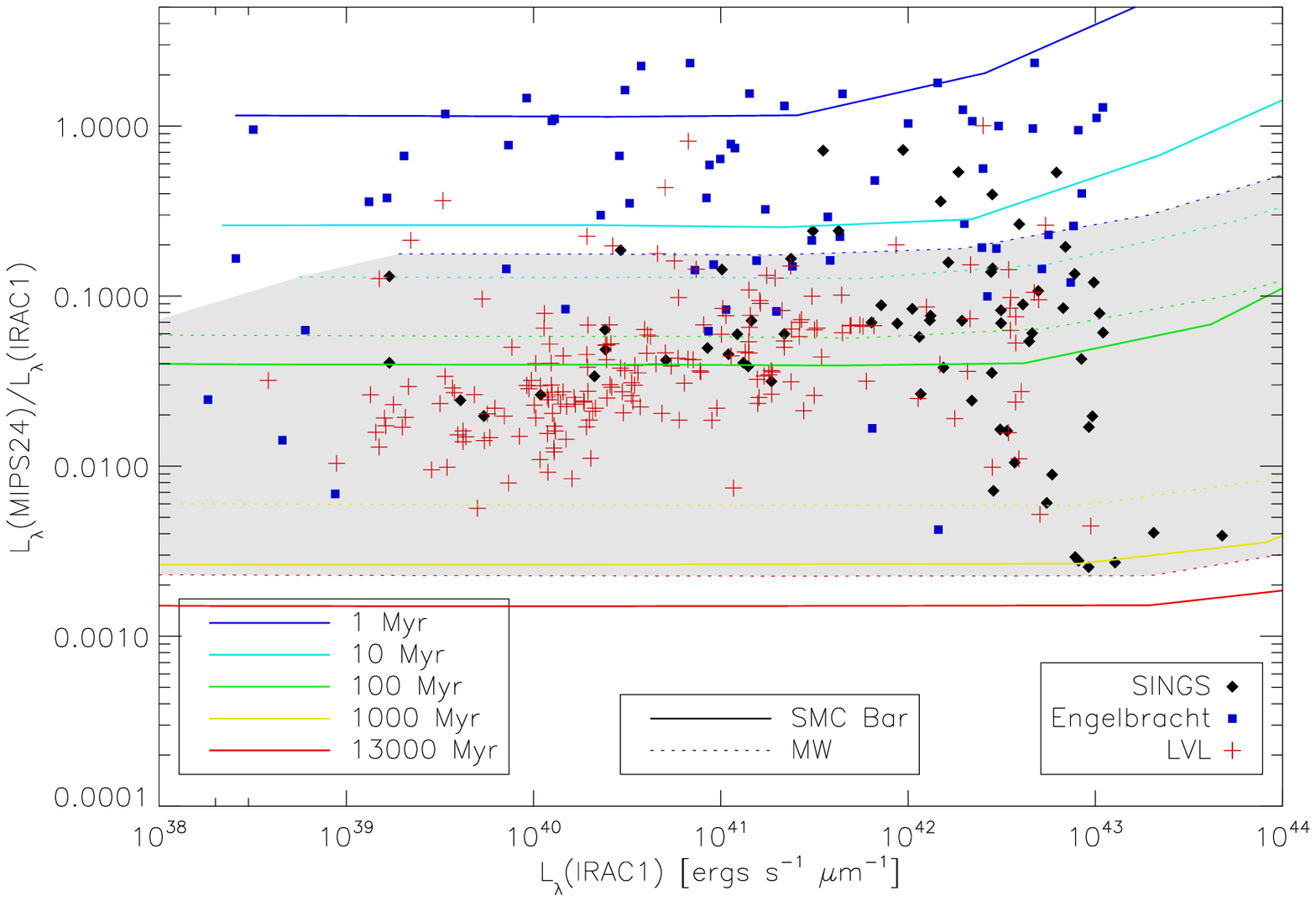}
	\plottwo{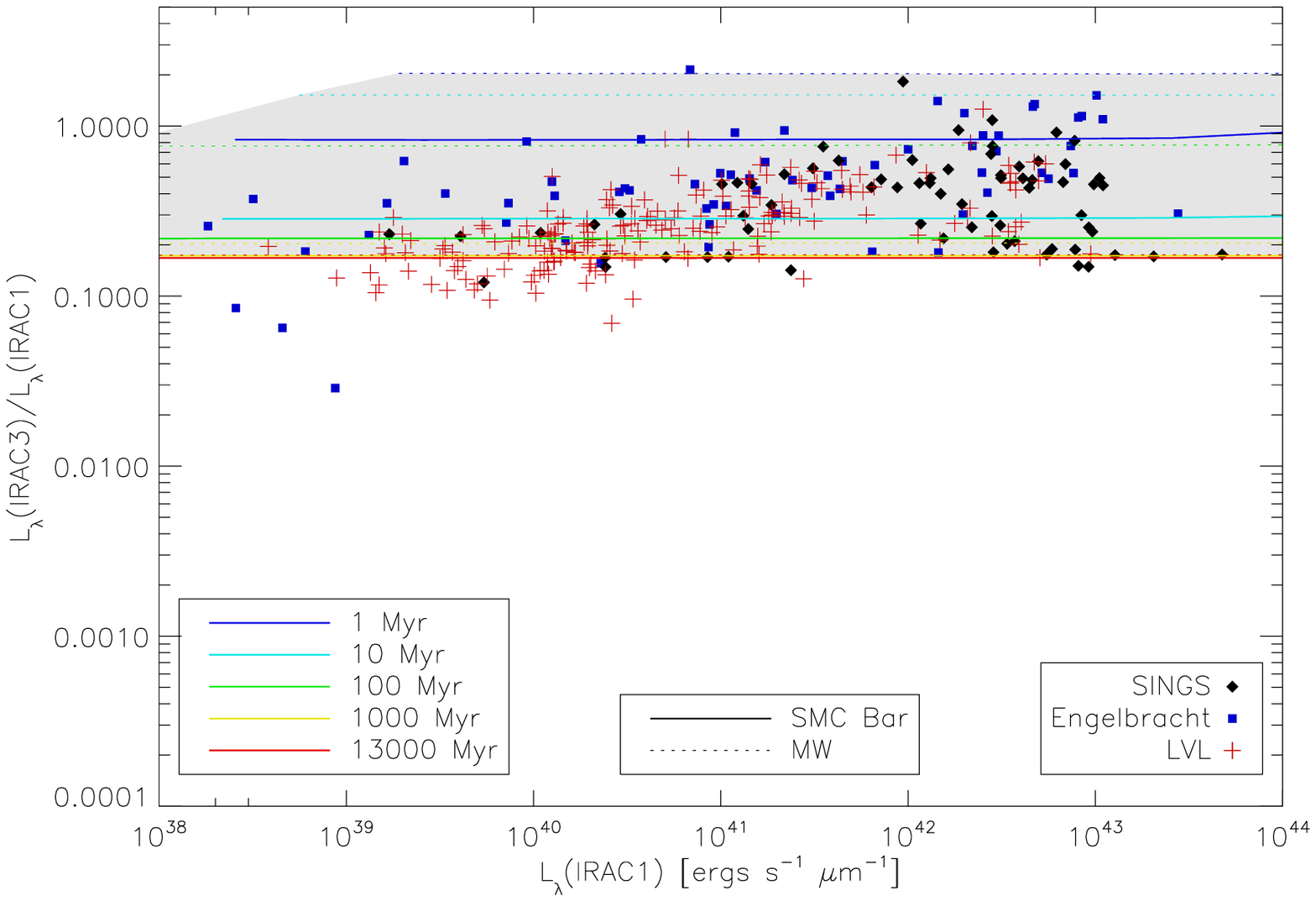}{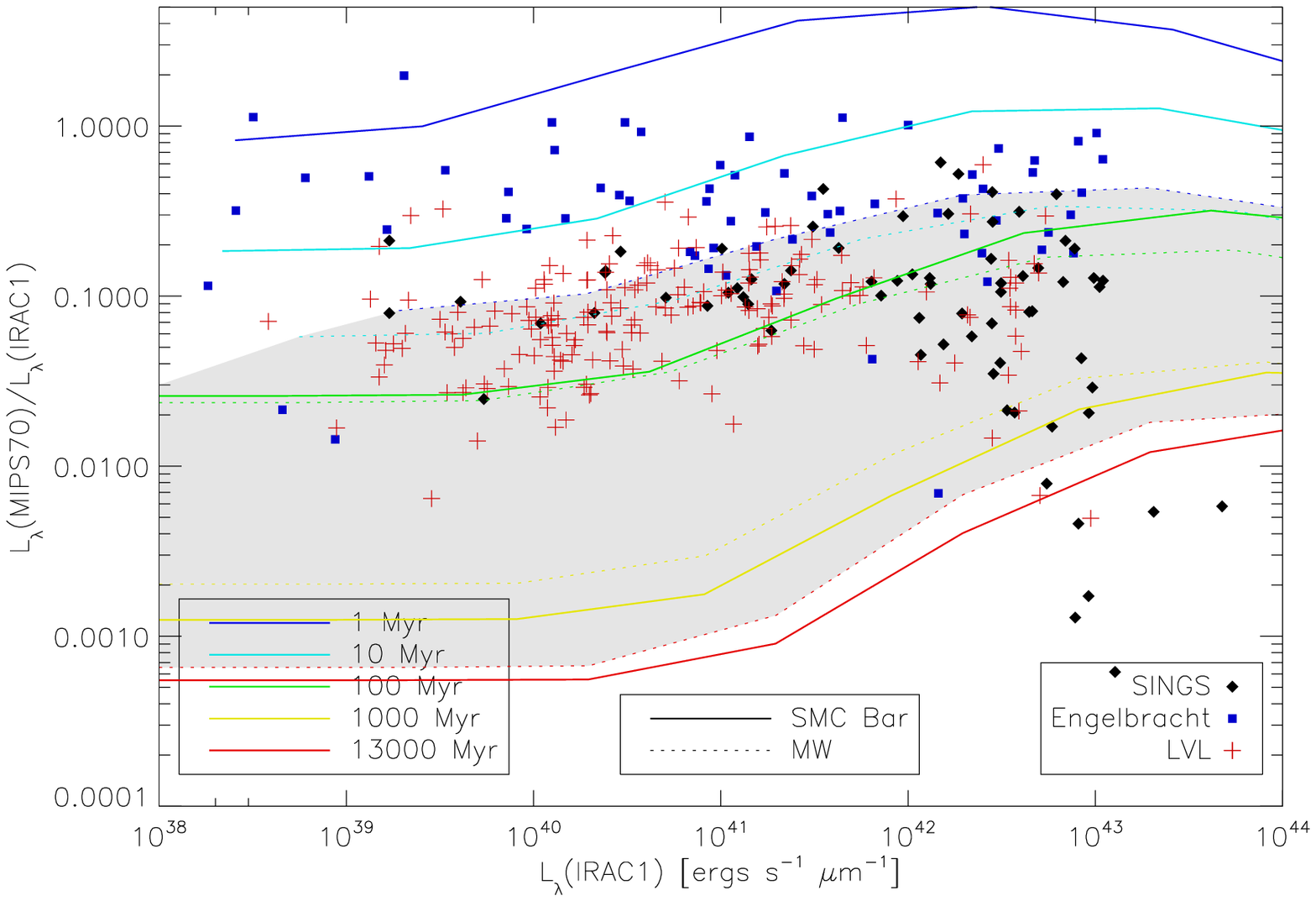}
	\plottwo{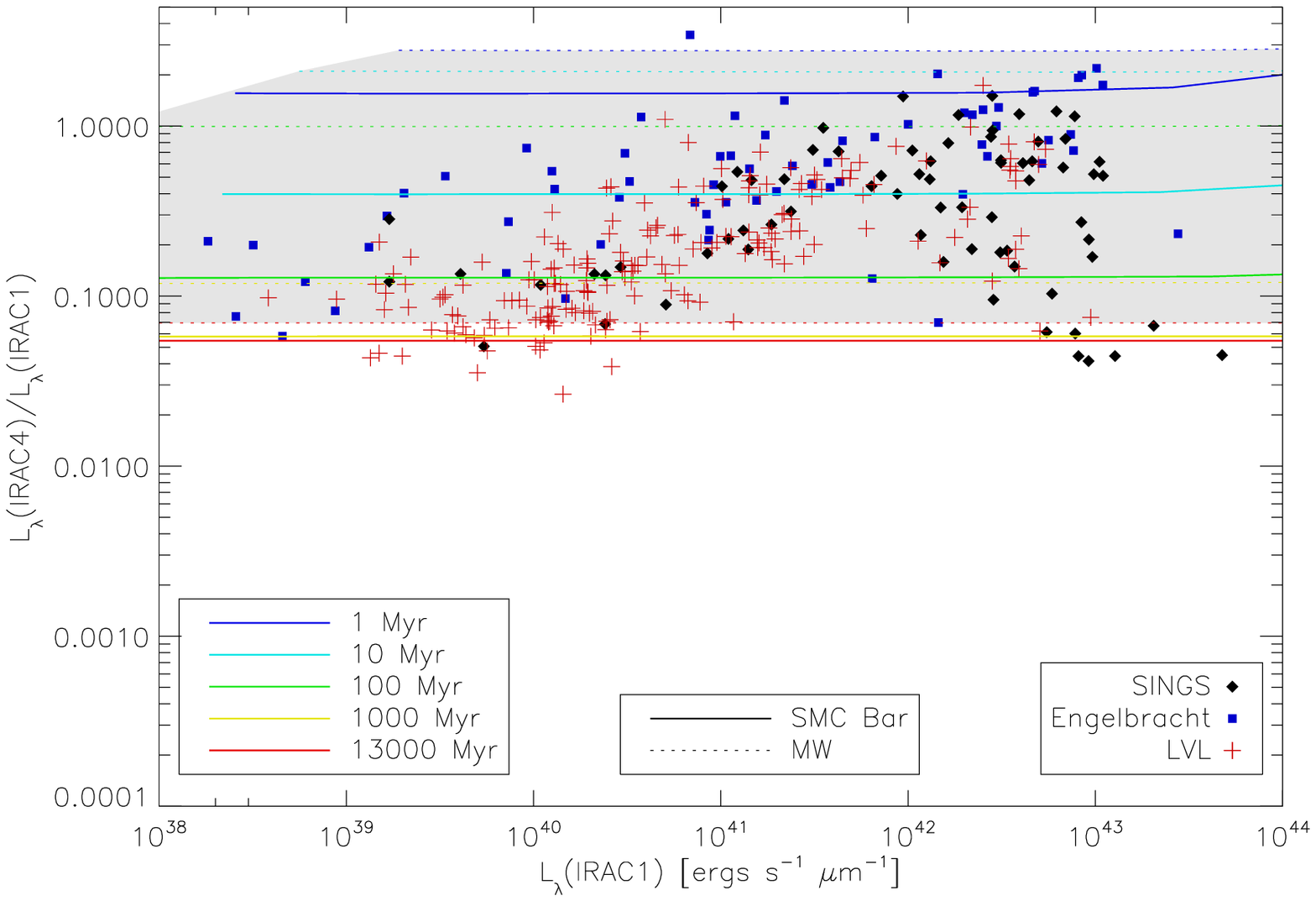}{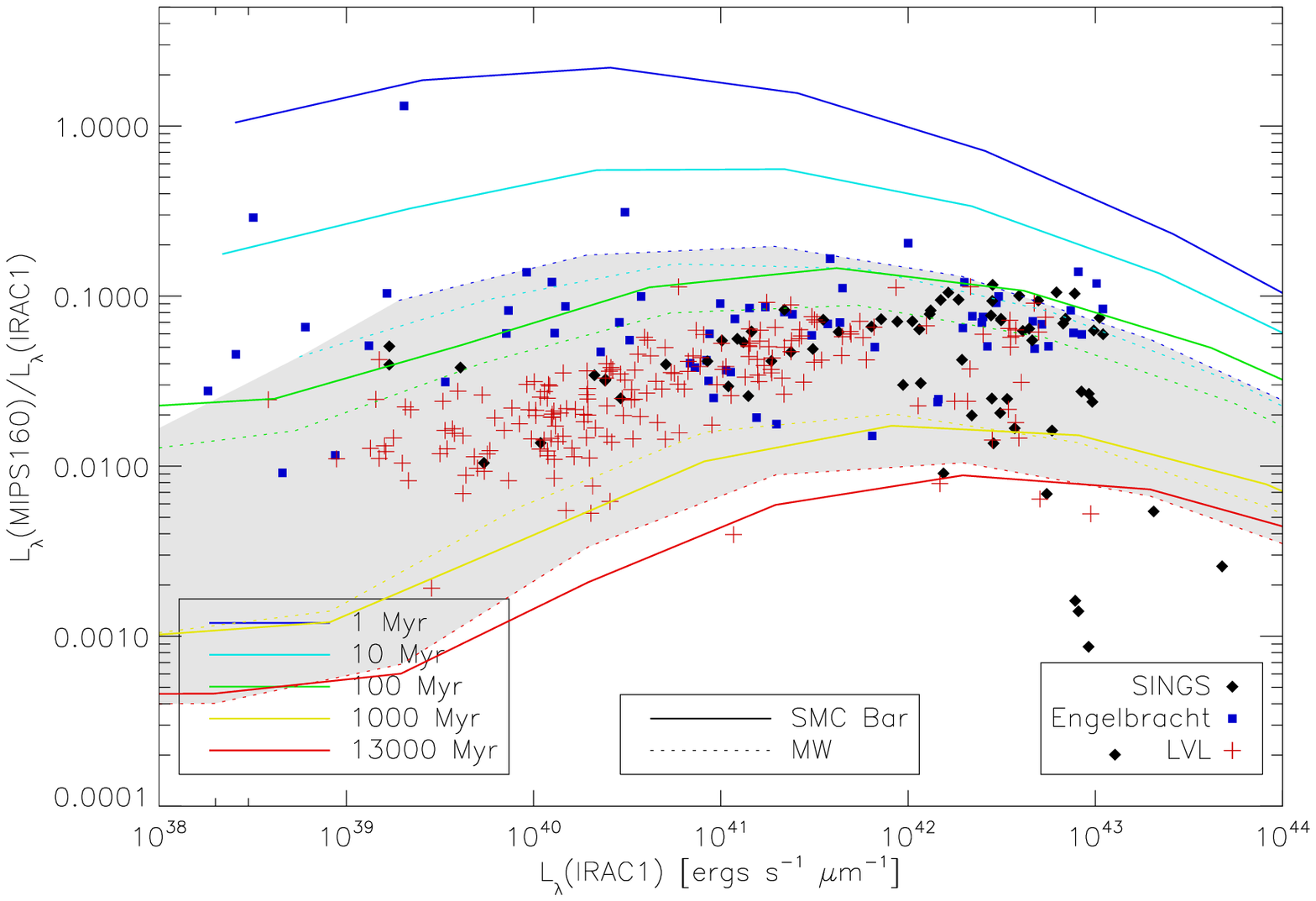}
	\caption{
		Luminosity ratios vs IRAC1 (3.6 \micron) luminosity plots for the
		IRAC (left) and MIPS (right) infrared bands. The models with different
		stellar ages (1 Myr to 13 Gyr old) are shown as curves with different
		colors; MW and SMC Bar type dust are in different line styles
		(dashed and solid respectively). The area covered by models with MW type dust
		is shaded in light gray, and is noticeably smaller then the range
		spanned by models with SMC Bar type dust in the far-IR. They are compared to these
		observed galaxies: the SINGS galaxies \citep{SINGS} in black
		diamonds, the \citet{Engelbracht2008} starburst galaxies
		in blue squares, and the LVL galaxies \citep{LVL} in red crosses.
		Here we use solar metallicity, $\tau_{v} = 1$ and a radius of 10 kpc.
	}
	\label{fig:panel}
\end{figure*}

In Figure~\ref{fig:panel}, we plot the luminosity ratios versus IRAC1
luminosity for IRAC and MIPS bands; the five curves correspond to
five different stellar ages.  The younger three (1, 10 and 100 Myr)
are modeled with the SHELL geometry and the older two (1 and 13 Gyr)
are modeled with the CLOUDY geometry.  Younger populations are more
efficient in producing IR due to being more embedded in the dust
and their much higher intrinsic L(UV)/L(IRAC1) ratios, and
the difference is most pronounced in the mid-IR.  Taking the
IRAC4/IRAC1 vs IRAC1 plot as an example, we see that the 10 Myr
line matches the data the best (among the models with SMC type dust).
The 1 Myr old models produce too much 8 \micron\ flux (per unit 3.6
\micron\ flux, on average), while the older models produce too little.
For the MIPS24/IRAC1 plot, a combination of the 10 Myr old and 100 Myr old
models match the median of the data, while for the MIPS160/IRAC1
plot, a combination of the 100 Myr old and 1 Gyr old models would do.
This shows that the age of the stellar populations that dominates IR emission is on the
order of 100 million years.  The difference from IRAC4 to MIPS160 shows
that the importance of younger stellar populations (less than 100 Myr) is relatively
higher at shorter wavelengths.

As we vary the age, the IRAC2-to-IRAC1 plot shows a different trend
from the other plots.  Both the data and the models fall into a very
narrow range (\td 0.5 dex).  This is because this color is dominated by stars only and
the colors of populations of stars in the mid-IR are relatively constant;
it is much less sensitive to dust compared to the longer wavelengths.

Note that the data points that have very low luminosity ratios in the
MIPS bands are elliptical galaxies.  The elliptical galaxies are shown as
red symbols in Figure~\ref{fig:sings}, and they show up in the lower right corner
in the MIPS70 and MIPS160 plots.  From our models, older stellar populations produce less
dust emission per unit near-IR luminosity, and therefore should be better
at explaining elliptical galaxies.  However, the luminosity ratios of these
galaxies are so low that even the oldest models (at about
the age of the universe) cannot explain them.  Noting that we have
used $\tau_{v} = 1.0$ and 10 kpc radius in Figure~\ref{fig:panel},
this result suggest that the optical depth of elliptical galaxies
could be lower than the average galaxy, or that their radii are larger
than our modeled value. The former is the most probable explanation,
as the zero-dust case for our oldest model gives a value of \td 0.0005 for
MIPS160/IRAC1,
low enough to explain the elliptical galaxies.

\subsection{Optical depth and radius}\label{tvradius}

\begin{figure*}
	\plottwo{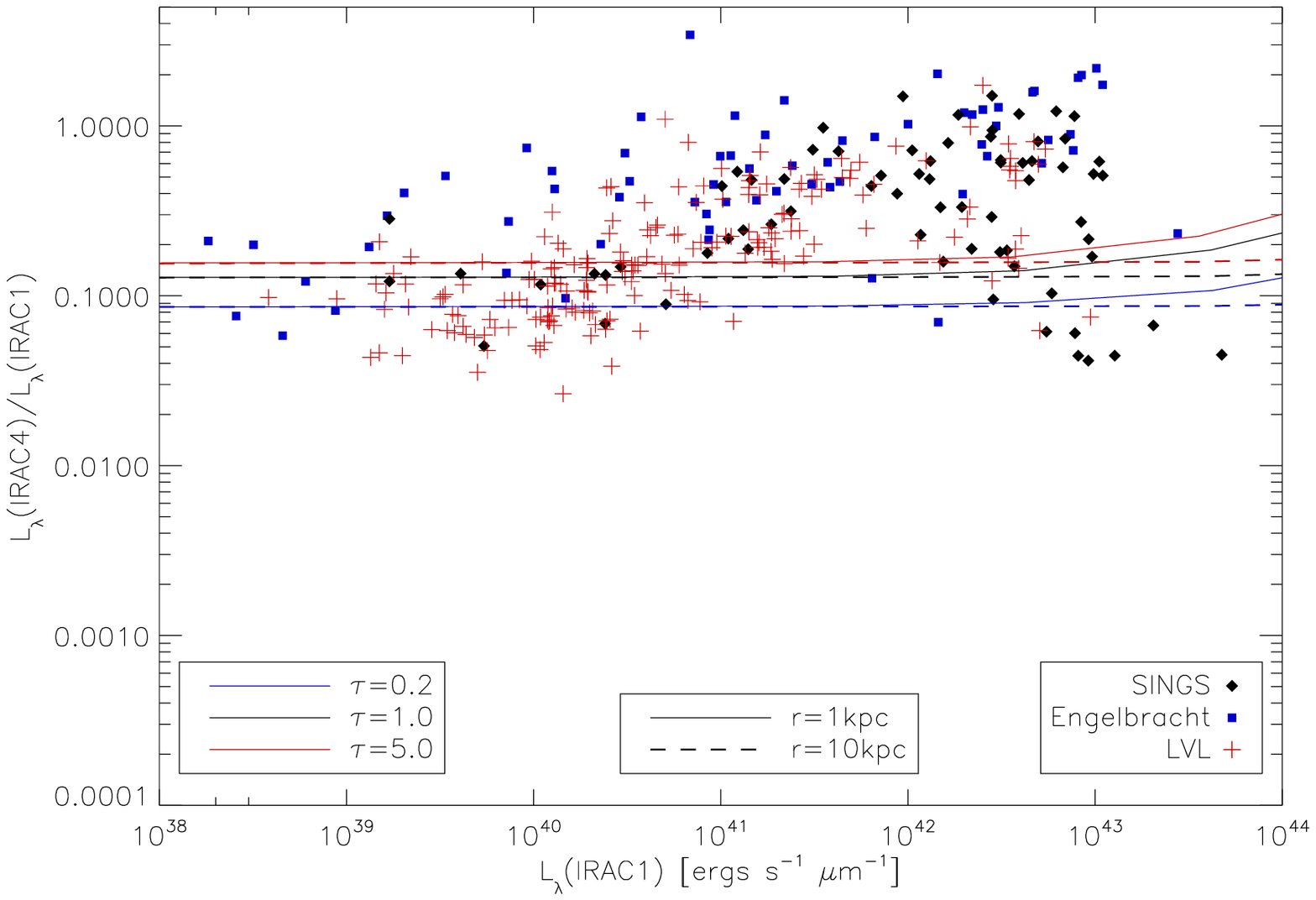}{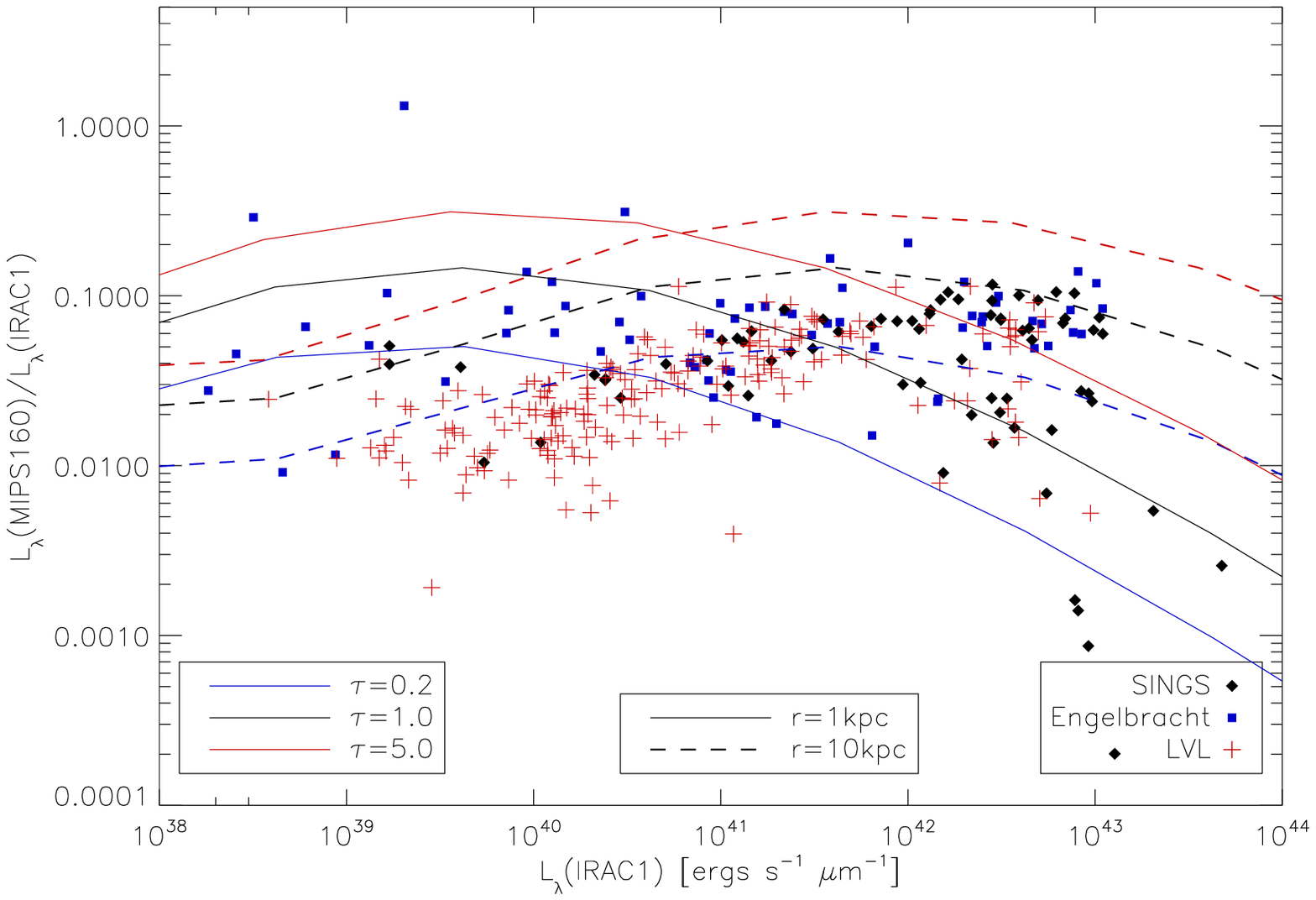}
	\caption{
		IRAC4 (8.0 \micron; left) and MIPS160 (160 \micron; right) to IRAC1 (3.6 \micron)
		luminosity ratios against IRAC1 luminosity for a fixed stellar age
		(100 Myr). Here we can study the effect of radius and optical depth
		in our model.
	}
	\label{fig:tauradius}
\end{figure*}

To illustrate the effect of the optical depth and radius of the model
region on the results,
we fix the stellar age at 100 Myr and vary the optical depth and radius
in Figure \ref{fig:tauradius}.  Dust with higher optical depth absorbs
more energy and emits more far-IR radiation, and therefore gives
a higher ratio in the plots.  A larger radius dilutes the
radiation field, lowers the dust temperature and shifts the equilibrium
dust emission peak to a longer wavelength in the SED.  In our luminosity
ratio diagrams, it shifts the curves horizontally to the right.
Because of the behavior of our model as explained in Section \ref{model},
we may also interpret the effect with mass surface density ($M/\pi r^2$).
With the same reasoning, as the mass surface density decreases, the curves
shift to the right.

As we study the effects of variations in optical depth and radius, we find a noticeable
difference for the IRAC bands and the MIPS bands.  For MIPS bands,
changing the two parameters can change the curves significantly.
This is evident on the plot for MIPS160 (Figure~\ref{fig:tauradius},
right).  However, in the IRAC bands, the curves are flatter and span a
narrower range.  As explained in section \ref{dust},
this is because the dust emission in the IRAC bands is dominated by
non-equilibrium heating.  And since the curves are flat, shifting
them horizontally makes no difference in our results; they are not
very sensitive to the choice of radius.  See section \ref{noneq} for a
discussion on equilibrium vs non-equilibrium dust emission.

From Figure~\ref{fig:tauradius} we can see that old stellar populations could possibly reproduce the observed
luminosity ratios in Figure \ref{fig:panel} only if the optical depth is
very high, say 5 or 10.  This is not realistic, as \citet{Holwerda2007}
has shown that for normal disk galaxies \tv~is on the order of unity.
In addition, if \tv~has such a high value, the optical depth at shorter
wavelengths will be even greater and we wouldn't be able to observe
these galaxies in UV.

\subsection{Equilibrium and non-equilibrium emission}\label{noneq}

\begin{figure*}
	\plottwo{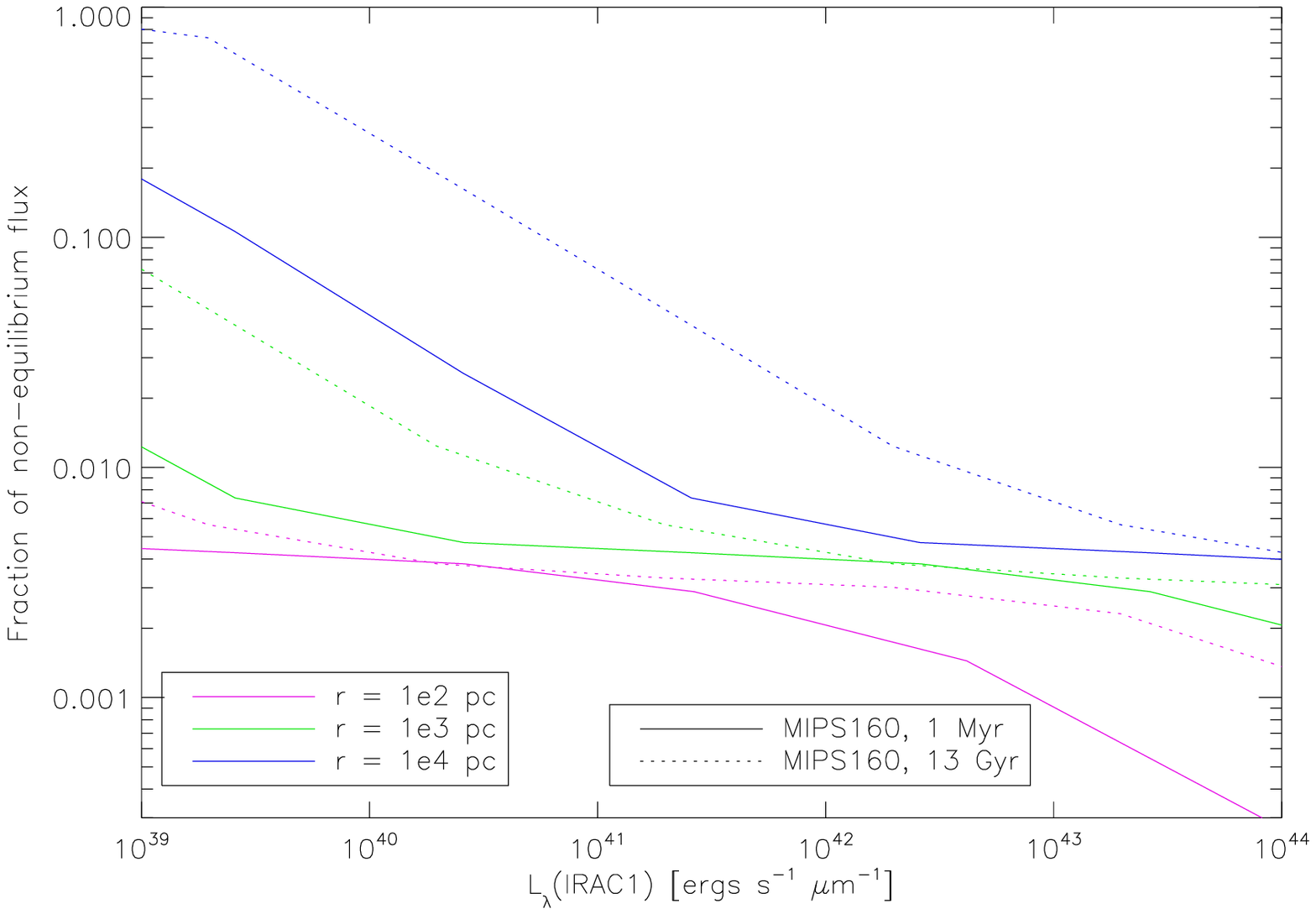}{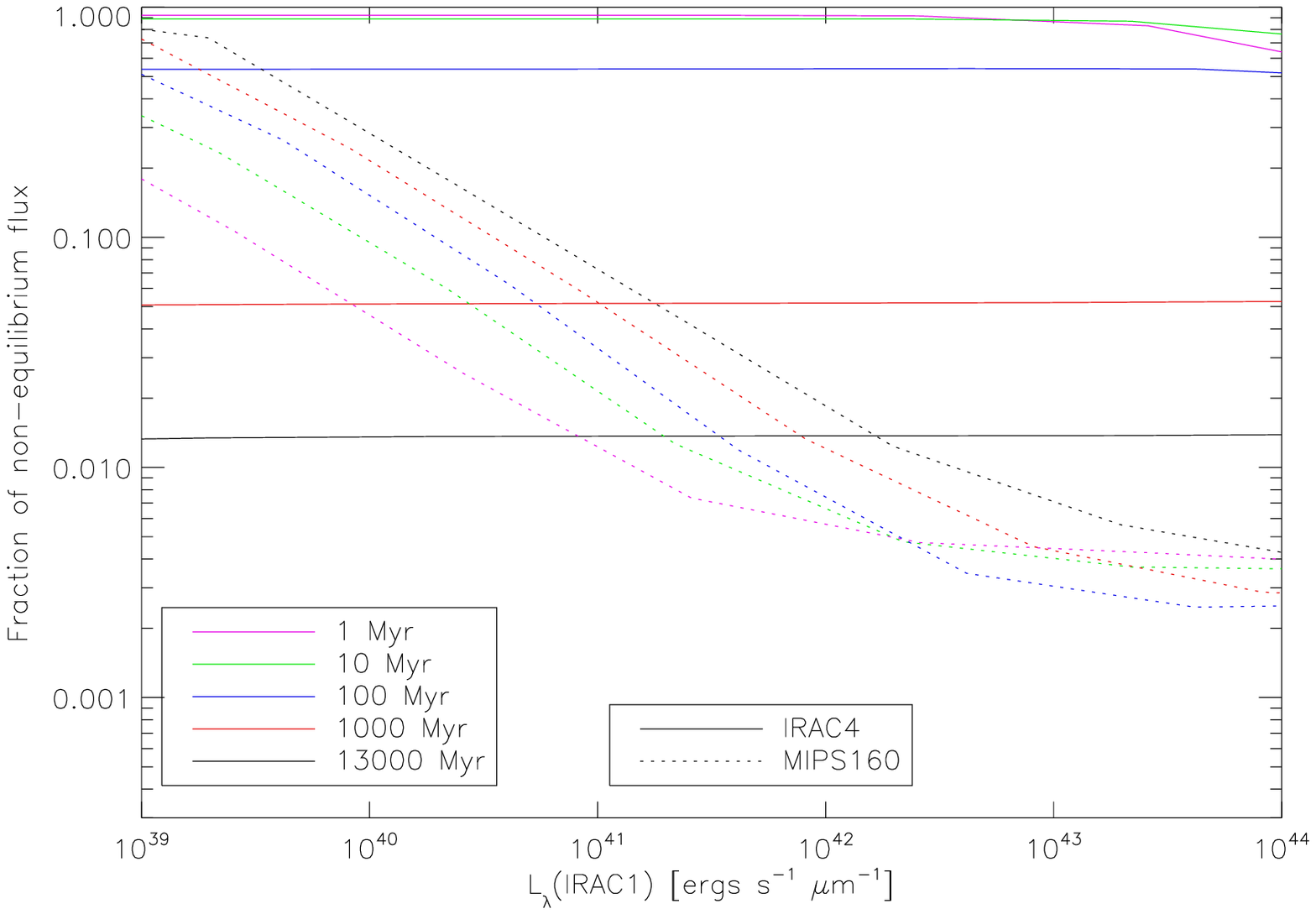}
	\plottwo{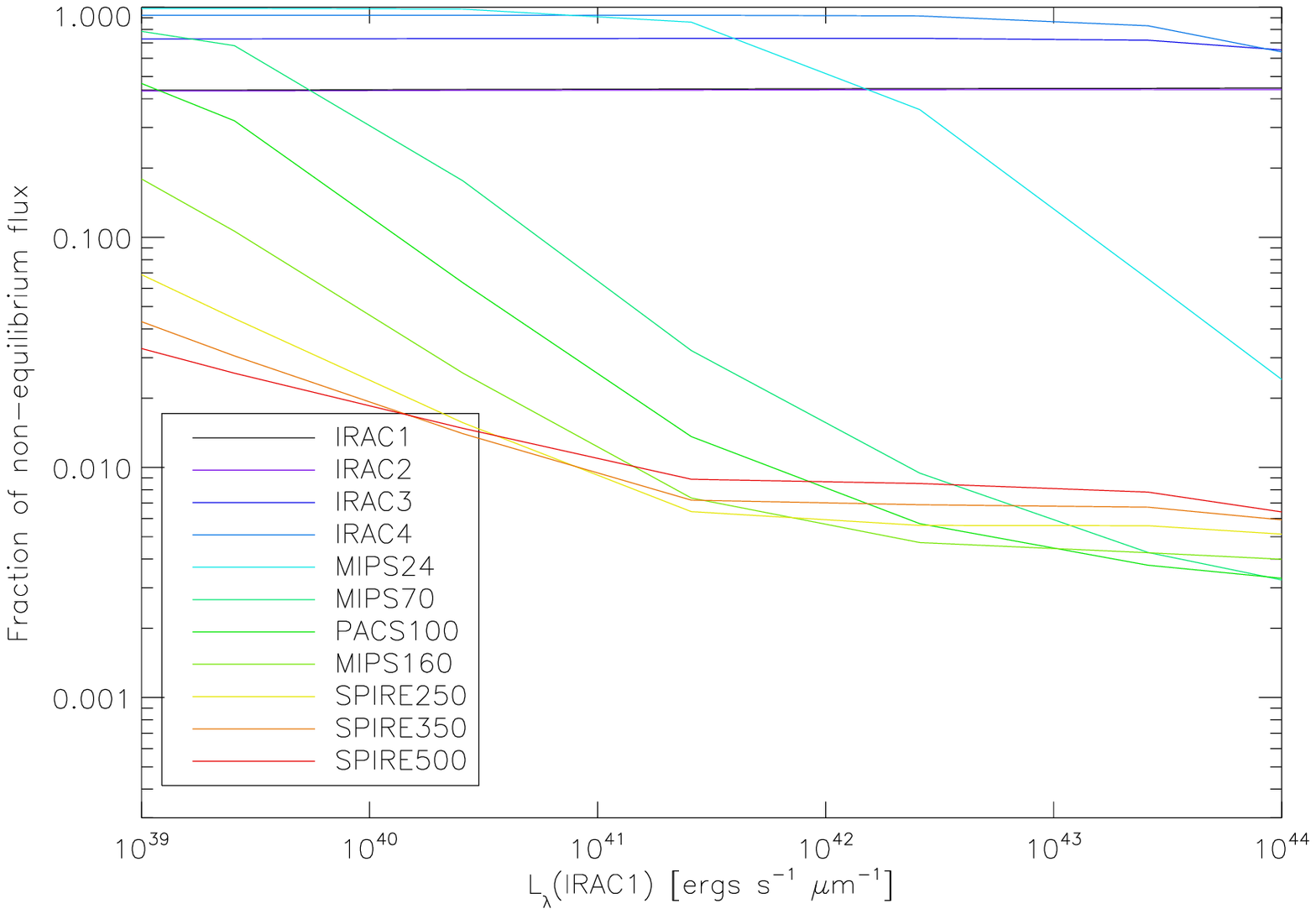}{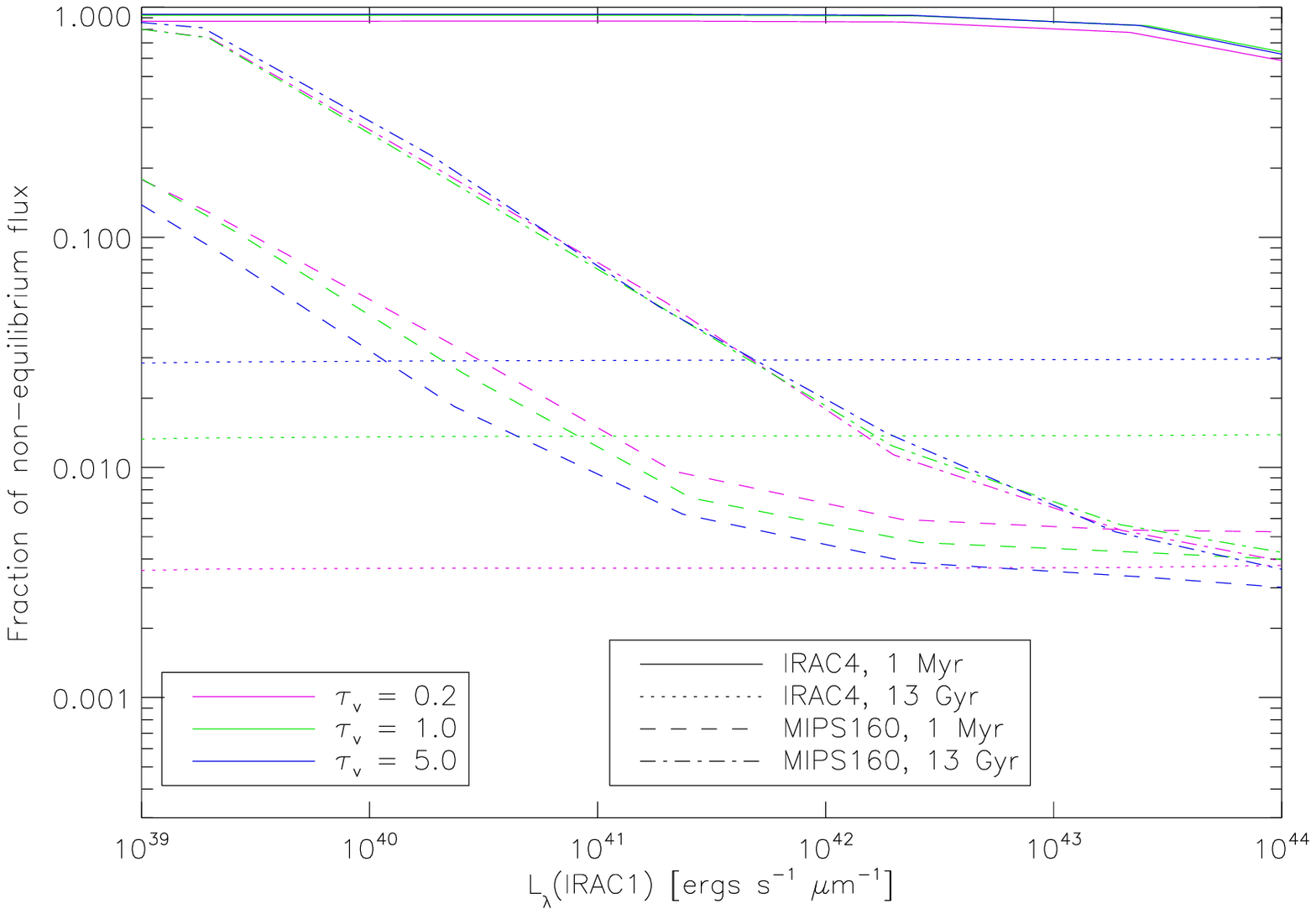}
	\caption{
		The modeled fraction of luminosity due to non-equilibrium heating plotted
		against the IRAC1 band luminosity for various model parameters.
		Unless specified in the legend, the model parameters are \tv~= 1, age = 1 Myr, and r = 10 kpc.
	}
	\label{fig:noneq}
\end{figure*}

To help interpret our results, we examine the fraction of luminosity due
to non-equilibrium heating.  The plots in Figure~\ref{fig:noneq}
shows the different behavior of the fraction in different IR bands.

In the far-IR (MIPS70 to SPIRE500), non-equilibrium
emission is negligible compared to equilibrium emission for a
wide range of IRAC1 luminosity.  Only at very low IRAC1 luminosity and
large ($>10$ kpc) radii does non-equilibrium emission contribute significantly in
the far-IR.  When the flux is dominated by equilibrium emission, it depends
on the dust temperature and therefore model parameters like the radius,
and this can be seen in the MIPS160 curves of Figure~\ref{fig:tauradius}.

On the other hand, the mid-IR bands (e.g. IRAC) exhibit different behavior.  When modeled with
young stellar populations (1 - 100 Myr old), the flux is dominated by non-equilibrium
emission because of the large amount of highly energetic UV photons.
When modeled with older stellar populations, the fraction of non-equilibrium emission
is small, so equilibrium emission is the major source of the IR flux
from dust. However, the dust temperature is not high enough to emit
significant energy in mid-IR, so the flux is dominated by the stellar
continuum.  Either way, equilibrium emission does not play a significant
role in the mid-IR bands, except at extremely high luminosities.
Therefore, the radius parameter (which affects the dust temperature)
does not change the result as much as they do for the far-IR bands.

Non-equilibrium emission is dominated by very high energy UV photons.
As opacity goes up with the energy of the photon, most high energy UV
photons are likely to be absorbed even in a low $\tau_{v}$ environment.
IRAC4 is dominated by non-equilibrium emission, therefore
further increases in $\tau_{v}$ do not result in more high energy UV photons
being absorbed and so the IRAC4 to IRAC1 luminosity ratio would not change significantly.
If we go back to Figure~\ref{fig:tauradius},
we see that the luminosity ratio is less sensitive to the optical depth
in IRAC4 then in MIPS160. MIPS24 is the turnover point for the two behaviors.
This explains why the curves behave so differently in IRAC plots and
MIPS plots as we change the model parameters in the previous sections.

\subsection{Constraints on the fraction of luminosity from old stars}\label{constrain}

In this section, we attempt to calculate the fraction of IR luminosity
that could be due to old stars using simple assumptions.  Assume there
are two non-interacting populations of stars, one younger and one older.
When we look at models with the same IRAC1 luminosity as an observed
galaxy, if the observed luminosity (say in MIPS160) is in between
the younger (higher) and older (lower) luminosities, there exists a
fraction $x$ of old stars of which the combination of the two model
stellar populations reproduces the observed luminosity.  We can write the
luminosity of the $i^{th}$ galaxy at wavelength $\lambda$, $L_{i}(\lambda)$ as 

\begin{equation}
L_{i}(\lambda) = [1 - x_{i}(\lambda)] Y_{i}(\lambda) + x_{i}(\lambda) O_{i}(\lambda)
\label{eqn:init}
\end{equation}

\noindent
where $x_{i}$ is the fraction of old stars, and $Y_{i}$ and $O_{i}$ are the
luminosities of the young and old models (e.g. 1 Myr and 13 Gyr)
that have the same IRAC1 luminosity
as the $i^{th}$ galaxy, respectively.  The two terms represent the
contribution of luminosity from the two populations.
Eq. \ref{eqn:init} can be inverted to solve for $x_{i}$ to yield

\begin{equation}
x_{i}(\lambda) =
\frac{L_{i}(\lambda)-Y_{i}(\lambda)}{O_{i}(\lambda)-Y_{i}(\lambda)}
\label{eqn:x}
\end{equation}

When we
use the formula for galaxies that have a luminosity (per unit IRAC1
luminosity) lower than the one given by the old model, $x_i$ will be
greater than 1, which is not physical.  This corresponds to the case
where the real galaxy is less efficient than the old model in producing
dust emission. Such a situation can arise if, for example, the galaxy
has a lower average optical depth than that assumed in our model.
As a simplification we simply set $x_i = 1$ in these cases.  On the other hand, for
galaxies with the luminosity per unit IRAC1 luminosity higher than our
young model, $x_i$ will be negative and we set $x_i = 0$.  After solving
for the fraction of old stars $x_i$, we can calculate the fraction of
luminosity due to old stars $f_{i}$:

\begin{equation}
f_{i}(\lambda) = \frac{x_{i}(\lambda) O_{i}(\lambda)}{L_{i}(\lambda)}
\label{eqn:f}
\end{equation}

\begin{table*}
	\caption{
		Calculated fractions of old stars - SINGS sample.
	}
	\centering
	\begin{tabular}{c c c c c c}
		\hline \hline
		Dust Type & Flux band & Fraction of & Fraction of luminosity & Number of galaxies & Number of galaxies \\
		  &  & old stars $x$ & from old stars $f$ & with $x_i = 1$ & with $x_i = 0$ \\
		\hline
		MW & IRAC4 & 0.854 $\pm$ 0.018 & 0.285 $\pm$ 0.078 & 5 & 1 \\
		MW & MIPS24 & 0.611 $\pm$ 0.102 & 0.068 $\pm$ 0.017 & 1 & 9 \\
		MW & MIPS70 & 0.569 $\pm$ 0.110 & 0.145 $\pm$ 0.071 & 3 & 8 \\
		MW & MIPS160 & 0.572 $\pm$ 0.107 & 0.181 $\pm$ 0.062 & 4 & 7 \\
		SMC Bar & IRAC4 & 0.736 $\pm$ 0.055 & 0.220 $\pm$ 0.061 & 2 & 1 \\
		SMC Bar & MIPS24 & 0.942 $\pm$ 0.008 & 0.050 $\pm$ 0.007 & 1 & 1 \\
		SMC Bar & MIPS70 & 0.964 $\pm$ 0.001 & 0.120 $\pm$ 0.053 & 3 & 1 \\
		SMC Bar & MIPS160 & 0.931 $\pm$ 0.004 & 0.203 $\pm$ 0.055 & 3 & 1 \\
		\hline
	\end{tabular}
	\label{table:constrain}
\end{table*}

\begin{table*}
	\caption{
		Calculated fractions of old stars -
                \citet{Engelbracht2008} sample.
	}
	\centering
	\begin{tabular}{c c c c c c}
		\hline \hline
		Dust Type & Flux band & Fraction of & Fraction of luminosity & Number of galaxies & Number of galaxies \\
		  &  & old stars $x$ & from old stars $f$ & with $x_i = 1$ & with $x_i = 0$ \\
		\hline
		MW & IRAC4 & 0.758 $\pm$ 0.049 & 0.191 $\pm$ 0.058 & 1 & 1 \\
		MW & MIPS24 & 0.184 $\pm$ 0.101 & 0.021 $\pm$ 0.006 & 1 & 40 \\
		MW & MIPS70 & 0.174 $\pm$ 0.083 & 0.022 $\pm$ 0.014 & 1 & 36 \\
		MW & MIPS160 & 0.516 $\pm$ 0.105 & 0.088 $\pm$ 0.016 & 1 & 8 \\
		SMC Bar & IRAC4 & 0.590 $\pm$ 0.097 & 0.139 $\pm$ 0.040 & 1 & 6 \\
		SMC Bar & MIPS24 & 0.630 $\pm$ 0.118 & 0.017 $\pm$ 0.003 & 1 & 7 \\
		SMC Bar & MIPS70 & 0.830 $\pm$ 0.039 & 0.020 $\pm$ 0.005 & 1 & 2 \\
		SMC Bar & MIPS160 & 0.913 $\pm$ 0.013 & 0.097 $\pm$ 0.009 & 1 & 1 \\
		\hline
	\end{tabular}
	\label{table:constrain_engel}
\end{table*}

\begin{table*}
	\caption{
		Calculated fractions of old stars - LVL sample.
	}
	\centering
	\begin{tabular}{c c c c c c}
		\hline \hline
		Dust Type & Flux band & Fraction of & Fraction of luminosity & Number of galaxies & Number of galaxies \\
		  &  & old stars $x$ & from old stars $f$ & with $x_i = 1$ & with $x_i = 0$ \\
		\hline
		MW & IRAC4 & 0.950 $\pm$ 0.006 & 0.512 $\pm$ 0.102 & 35 & 1 \\
		MW & MIPS24 & 0.775 $\pm$ 0.050 & 0.081 $\pm$ 0.013 & 2 & 9 \\
		MW & MIPS70 & 0.512 $\pm$ 0.076 & 0.031 $\pm$ 0.012 & 2 & 25 \\
		MW & MIPS160 & 0.803 $\pm$ 0.029 & 0.155 $\pm$ 0.029 & 4 & 1 \\
		SMC Bar & IRAC4 & 0.887 $\pm$ 0.021 & 0.410 $\pm$ 0.083 & 16 & 1 \\
		SMC Bar & MIPS24 & 0.954 $\pm$ 0.005 & 0.061 $\pm$ 0.008 & 1 & 1 \\
		SMC Bar & MIPS70 & 0.945 $\pm$ 0.008 & 0.028 $\pm$ 0.009 & 2 & 1 \\
		SMC Bar & MIPS160 & 0.969 $\pm$ 0.003 & 0.128 $\pm$ 0.022 & 4 & 1 \\
		\hline
	\end{tabular}
	\label{table:constrain_lvl}
\end{table*}

Using 1 Myr (young) models and 13 Gyr (old) models, we calculate
the fractions $x$ and $f$ for the 3 samples of galaxies, and tabulate
the results for the IRAC4, MIPS24, MIPS70 and MIPS160 bands in
Tables~\ref{table:constrain}-\ref{table:constrain_lvl}.  Again, \tv = 1 and 10 kpc radius are
used.  The fractions are shown here as the average plus or minus the
standard deviation.  The number of galaxies with out-of-range $x_i$
in each sample is given in columns 4 and 5.  They are adjusted to $x_i$ = 0 or 1 as explained above.

While the fraction of old stars $x$ can exceed 90\% (as in some of the calculation
for models with SMC Bar dust), the fraction of the luminosity produced
by old stars, $f$, is much
lower; it is generally lower than 20\%, with the exception of the
IRAC4 band for the LVL galaxies.  Although $f$ is high for LVL/IRAC4,
the remaining bands in the LVL sample yield lower values of $f$, consistent with
the other 
two galaxy samples.  If we take the LVL/IRAC4 combination
out of the picture and restrict our results to SMC Bar dust,
the highest value of $f$ is 22.0\%.  Or, if we use 10 (100) Myr old stars instead
of 1 Myr old stars as the younger population, the highest fraction becomes
27.9\% (25.2\%), again with MW/IRAC4/SINGS.  On the other hand, if we keep the 1 Myr
old stars but change the older population to 1 Gyr old stars, the highest
fraction becomes 42.8\%.

The fraction of luminosity $f$ is generally higher for MW type dust,
but the choice of dust does not affect our conclusion that dust emission
is dominated by young stars.  It is remarkable that the values of $f$
computed from the MIPS luminosities are similar for both
types of dust even when the fractions of old stars $x$
is consistently higher for SMC Bar type dust (this is generally true for
different stellar ages as well).  
This could be attributed to the steeper far-UV rise in the SMC Bar
extinction curve; for the same value of \tv, the SMC Bar type dust is more
effective in absorbing far-UV photons, and therefore requires fewer young
stars to produce the same IR luminosity.  The higher number of galaxies with
$x_i = 0$ for MW type dust can be understood by looking at
Figure~\ref{fig:panel}; the MW 1 Myr old line is below a significant number
of galaxies.

If we do a more complex study to include more than two stellar ages in this
analysis, the fraction $f$ will only be lower.
We assume a linear contribution from the 2 stellar populations,
which is not necessarily accurate because the effect of dust temperature on
dust emission is non-linear. The more well-mixed the young and old stars are,
the more non-linear their contribution will be.
However, as older stars release many fewer UV photons
compared to younger stars, it is almost certain that increasing the fraction of 
old stars would decrease the far-IR/IRAC1 ratios. Therefore, the far-IR/IRAC1 ratios
are monotonically decreasing functions of the fraction of old stars (when
all other parameters are fixed). Real stellar populations include a
wide range of stellar ages, and with our analysis we can see whether
such populations are closer to the ``old" population or the ``young" population.
While this simple analysis cannot tell us what the best stellar age for the
observed galaxies is, and the calculated fraction of old stars may suffer from
non-linear effects, it is evident that dust emission is dominated by young stars.

\subsection{Metallicity}\label{metal}

\begin{figure*}
	\plottwo{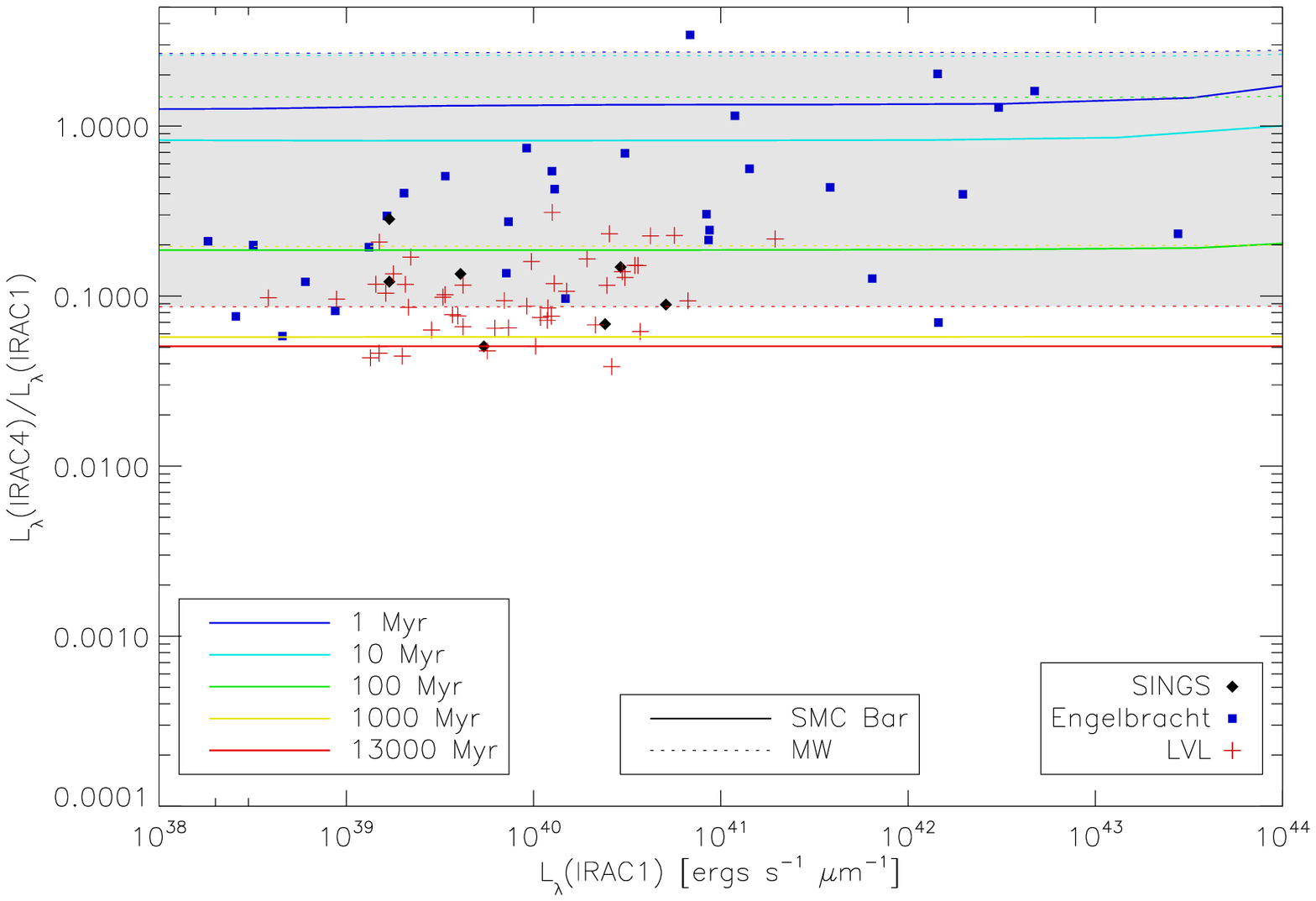}{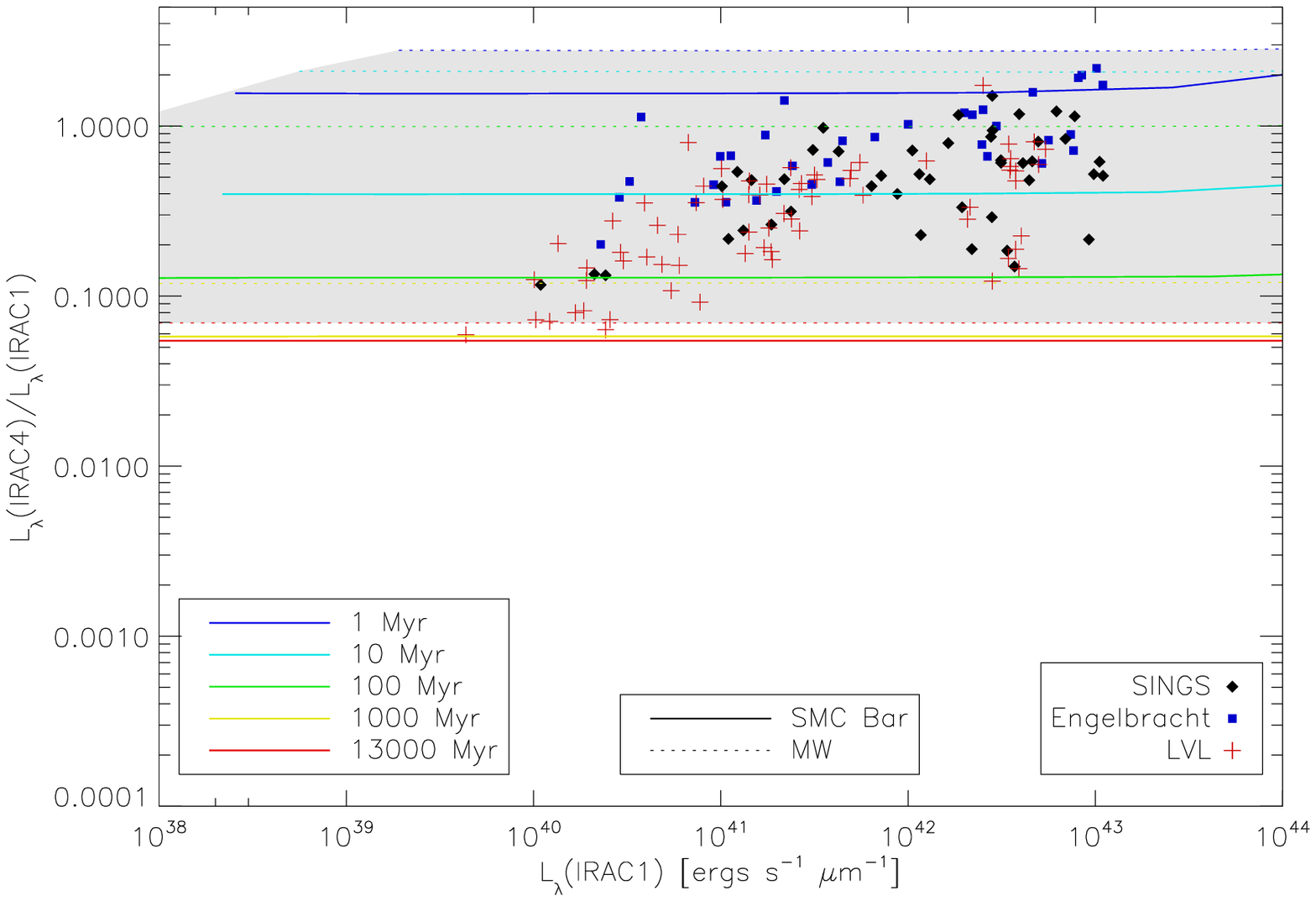}
	\plottwo{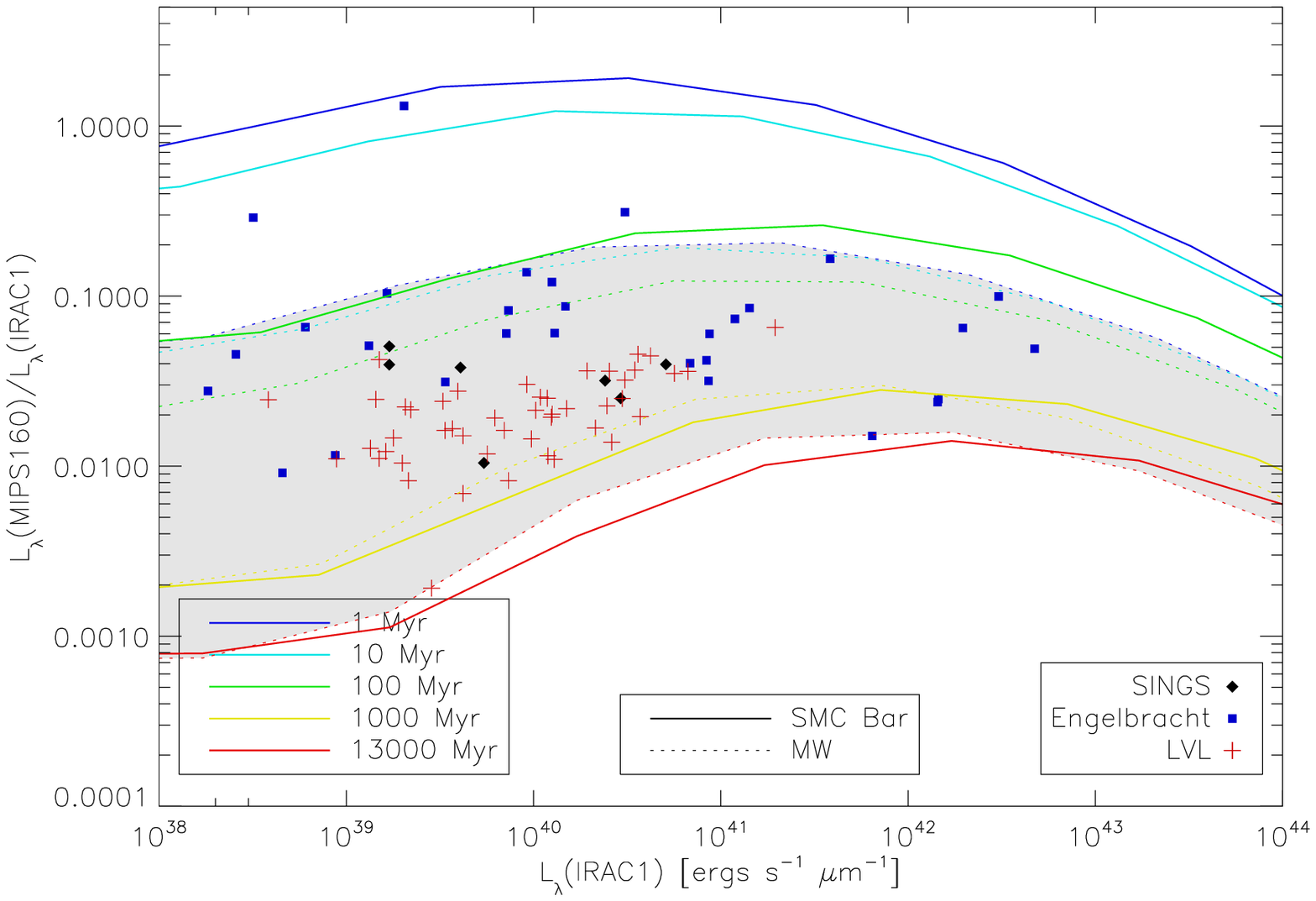}{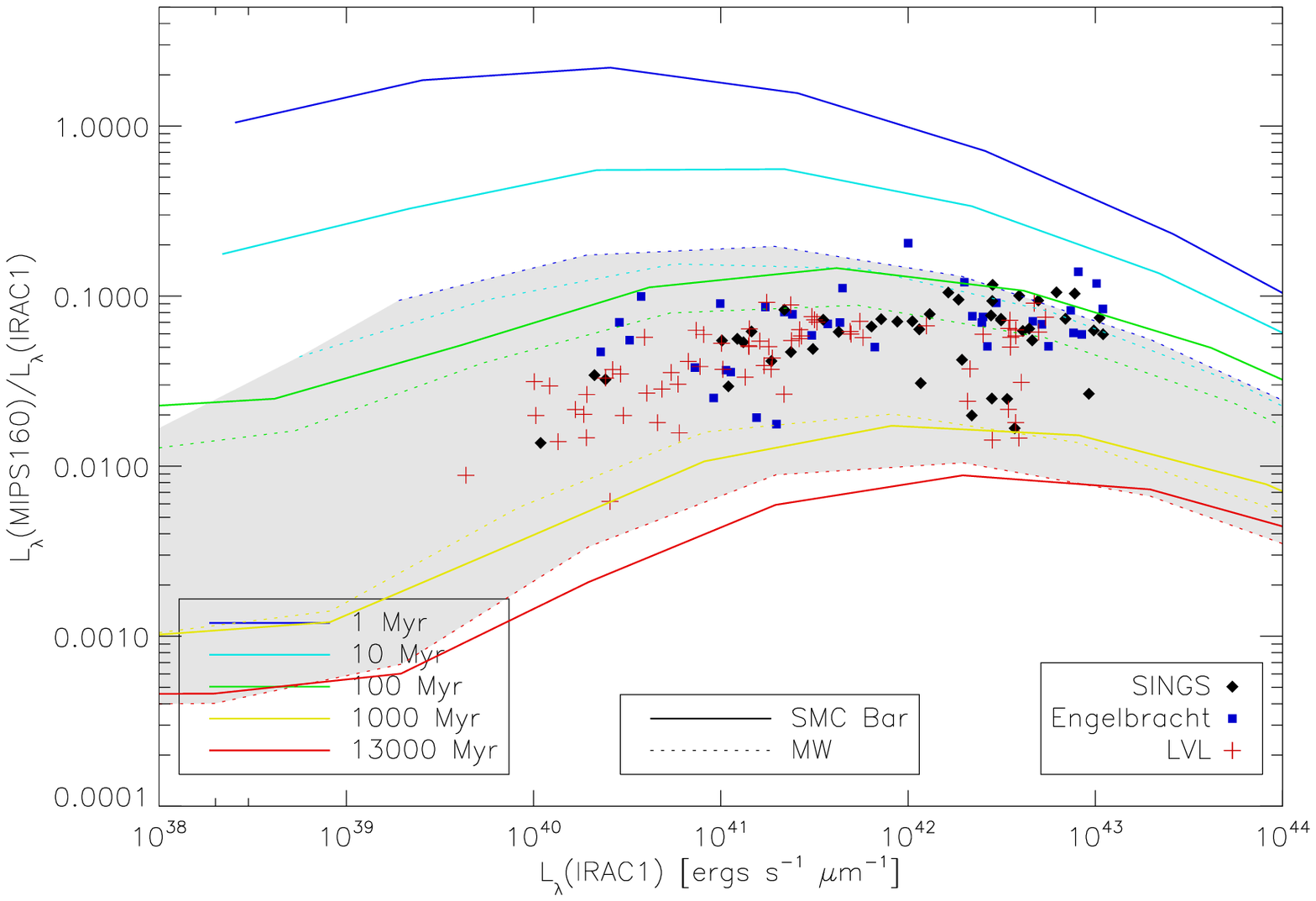}
	\caption{
		Plots to illustrate the effect of metallicity.
		The top (bottom) two figures are IRAC4/IRAC1 (MIPS160/IRAC1) luminosity ratio vs IRAC1 luminosity plots.
		Galaxies with metallicity lower than 8.1 are plotted with 1/5 solar metallicity models on the left,
		while the high metallicity galaxies are plotted with solar metallicity models on the right.
	}
	\label{fig:metal}
\end{figure*}

In stars with high metallicity, absorption lines form a continuum that
leads to the production of a soft UV radiation field.
This stellar atmosphere effect is known as ``line blanketing".
In contrast, stars with lower metallicity produce a harder radiation field.
Empirically, metallicity correlates with the amount of dust
(and therefore a low metallicity is often associated with a high gas-to-dust ratio).
In the previous sections, we use a fixed (solar) metallicity for our models.
However, the combination of a harder radiation field and a lower optical depth may give different results.

Metallicity has a somewhat phase-transition like effect.
For the relatively high metallicities observed in the LMC and the Milky Way,
observables are qualitatively similar.
However, when we go to the low metallicity found in SMC,
we find qualitatively different results, such as the lack of aromatic emission.
For example, \citet{Engelbracht2005} found an abrupt change in the 8-to-24 \micron\ color at around 1/4 solar metallicity.
\citet{Calzetti2010} choose a metallicity of $log(O/H)=8.1$ as a rough dividing line for the two behaviors.
We use the same value to divide our sample into two groups:
the high metallicity galaxies with metallicity higher than 8.1,
and the low metallicity galaxies with metallicity lower than 8.1.

In Fig.~\ref{fig:metal}, we compare the high metallicity galaxies with the low metallicity ones.
We plot the high metallicity galaxies with solar metallicity models (as in Fig.~\ref{fig:panel}),
but 1/5 solar metallicity models for the low metallicity galaxies.
The low metallicity sample has lower luminosity ratios in both IRAC4 and MIPS160 bands,
showing that metallicity does have an effect on dust heating.
They also have lower IRAC1 luminosity on average,
consistent with the view that the big luminous galaxies are more evolved and have more metals,
while the less luminous galaxies have relatively more young stars and lower metallicity.
Calculations of the fraction of luminosity from old stars ($f$) shows that
the results for the MIPS bands are qualitatively the same for the two metallicity groups;
but for the IRAC4 band, $f$ of the low metallicity group can be as
high as two times that of the high metallicity group.
For SINGS and LVL, $f$ of the low metallicity group is higher than 0.5 for IRAC4.
It suggests that the non-starburst, low metallicity galaxies may have lower PAH abundance than our dust model.
Since there is no significant trend in the MIPS bands, metallicity
does not affect our main conclusion. 

\subsection{Sub-mm Predictions}\label{herschel}

\begin{table*}
	\caption{
		Fractions of luminosity due to old stars at 100, 250, 350, and 500 \micron.
	}
	\centering
	\begin{tabular}{c c c c c c}
		\hline \hline
		Data Sample & Dust Type & $f(100 \micron)$ & $f(250 \micron)$ & $f(350 \micron)$ & $f(500 \micron)$ \\
		\hline
		Engelbracht & MW & 0.046 $\pm$ 0.006 & 0.136 $\pm$ 0.028 & 0.176 $\pm$ 0.036 & 0.213 $\pm$ 0.044 \\
		Engelbracht & SMC Bar & 0.032 $\pm$ 0.001 & 0.227 $\pm$ 0.028 & 0.349 $\pm$ 0.039 & 0.463 $\pm$ 0.044 \\
		SINGS & MW & 0.129 $\pm$ 0.059 & 0.237 $\pm$ 0.067 & 0.277 $\pm$ 0.072 & 0.308 $\pm$ 0.077 \\
		SINGS & SMC Bar & 0.111 $\pm$ 0.052 & 0.355 $\pm$ 0.053 & 0.477 $\pm$ 0.049 & 0.572 $\pm$ 0.045 \\
		LVL & MW & 0.105 $\pm$ 0.025 & 0.311 $\pm$ 0.037 & 0.397 $\pm$ 0.040 & 0.463 $\pm$ 0.042 \\
		LVL & SMC Bar & 0.068 $\pm$ 0.021 & 0.375 $\pm$ 0.029 & 0.566 $\pm$ 0.025 & 0.706 $\pm$ 0.020 \\
		\hline
	\end{tabular}
	\label{table:herschel}
\end{table*}

The Herschel Space Observatory was recently launched in May 2009.
The far infrared imaging camera of the Spectral and Photometric
Imaging Receiver (SPIRE) has 3 photometric bands centered at 250, 350
and 500 \micron.  In addition, the Photodetector Array Camera and
Spectrometer (PACS) has a photometric band at 100 \micron\ that MIPS
does not have. It is interesting to see, from a model point of view,
the contribution of luminosity of old stars at these wavelengths.

Table~\ref{table:herschel} shows the fraction of luminosity ($f$) due
to old stars at 100, 250, 350 and 500 \micron,
calculated with Equation~\ref{eqn:herschel}.  For the $i$th galaxy, we use
the fraction of old stars ($x_i$) calculated with Equation~\ref{eqn:x}
for MIPS160 and the luminosity from the corresponding
old ($O_i$) and young models ($Y_i$) to estimate $f_i$.
The denominator is the estimated luminosity at the SPIRE wavelengths,
and the numerator is the luminosity due to old stars only.

\begin{equation}
f_{i}(\lambda) =
\frac{x_i(160 \micron) O_i(\lambda)}{x_i(160 \micron) O_i(\lambda) + (1 - x_i(160 \micron)) Y_i(\lambda)}
\label{eqn:herschel}
\end{equation}

We convolve our model SEDs with the relative spectral response functions
of the SPIRE bands (SPIRE Observers' Manual, 2010) and the PACS
100 \micron\ band (PACS Observer's Manual, 2010) to compute 
the band integrated luminosity.  We calculate the average and
standard deviation for each combination of data sample (SINGS,
Engelbracht and LVL) and model dust type (MW, SMC bar).  We choose
MIPS160 for the estimation of $x$ because it is the longest wavelength
in this study.  As we go from 100 \micron\ to 500 \micron, $f$
increases, showing that old stars are increasing in importance at longer
wavelengths.

Moreover, $f$ in general has an increasing trend from the Engelbracht to the
SINGS and the LVL sample. We have similar trends in some of the other bands
but the trend in the SPIRE bands is much clearer, especially in the longest
wavelength. This shows the order of the importance of the old stars in
the 250-500 \micron\ regime for the three catalogs.
It is easy to understand the lower $f$
in the starburst sample as it has more recent star formation activity.
So for starburst galaxies, young stars still contribute more luminosity 
in the sub-mm range (up to 500 \micron) compared to old stars.
The difference between $f$ of LVL and $f$ of SINGS is likely due to a
combination of their stellar populations, galaxy type and composition.
We will continue to explore this wavelength regime and have a better
understanding when Herschel data is available.

\section{CONCLUSION}\label{conclusion}

Using our dusty radiative transfer model, together with IRAC and 
MIPS observations on the SINGS galaxies, the starburst galaxies
in \citet{Engelbracht2008} and the LVL galaxies, we have studied the effect
of stellar age on infrared luminosity.  We found that MW type dust tends to
produce too little infrared luminosity for some of the galaxies
(especially for the starburst galaxies), and so
SMC Bar type dust is a more appropriate choice.  However, we also note
that we have similar results with both types of dusts when we calculate
the luminosity due to old stars. 

From an analysis of the IRAC4/IRAC1 and MIPS160/IRAC1 luminosity
ratios vs IRAC1 luminosity plots,
we found that the observed luminosity cannot be produced 
by 13 Gyr old stellar populations alone.  The  stellar age that dominates dust
heating is on the order of 100 Myr.  However, a small number of
galaxies - the elliptical galaxies - did not fit well into our analysis.  
Their lower far-IR to IRAC1 ratios could be attributed to their deficiency
of dust.

We found that the models are more sensitive to changes in parameters
(such as stellar age, radius and optical depth) in
the far-IR bands compared to the mid-IR bands.  This can be explained
by the fact that non-equilibrium emission dominates mid-IR, but is
mostly negligible in far-IR.  When non-equilibrium emission dominates,
the luminosity ratio is less dependent on dust temperature and is
therefore less affected by changes in the radius and stellar mass;
it is also less sensitive to optical depth $\tau_{v}$ because the
extinction for highly energetic photons saturates.

With the simplistic assumption that the observed galaxies are composed
of two stellar populations of different ages, we found that the fraction
of far-IR luminosity from 13 Gyr old stars is generally less than 20\%.
The result does not depend on the metallicity.
Therefore, cold does not necessarily mean old; our study shows that far-IR
radiation is dominated by a small number of younger stars.

We are currently building a large grid of dusty radiation models and spectral
evolutionary synthesis models.  With the model grid we will attempt to
further confirm this study by fitting each galaxy individually and derive
properties such as stellar age. We will compare the statistics of the
resultant properties to the results in this paper and explain any
differences or new features observed.

\newpage
\bibliographystyle{astroads-karl}

\end{document}